\documentclass[a4paper,12pt]{article}
\usepackage{graphicx}% Include figure files
\usepackage[dvipsnames, svgnames]{xcolor}
\usepackage{dcolumn}% Align table columns on decimal point
\usepackage{bm}% bold math
\usepackage[numbers, sort&compress]{natbib}
\usepackage{amsmath, amssymb}
\usepackage{amsthm}
\usepackage{lipsum}
\usepackage[colorlinks, citecolor=blue, linkcolor=NavyBlue]{hyperref}
\usepackage{authblk}
\usepackage{float}
\usepackage{caption}
\usepackage{grffile}
\usepackage[title]{appendix}

\usepackage{lineno}

\begin{document}

\title{A new approach to handling factorial moment correlations through principal component analysis}

\author[1]{N.~Davis\footnote{email: nikolaos.davis@ifj.edu.pl}}
\affil[1]{Institute of Nuclear Physics PAN, ul. Radzikowskiego 152, 31-342, Krak\'ow, Ma\l{}opolska, Poland}
\date{}

\maketitle

\begin{abstract}
Intermittency analysis of factorial moments is a promising method used for the detection of power-law scaling in high-energy collision data. In particular, it has been employed in the search of fluctuations characteristic of the critical point (CP) of strongly interacting matter. However, intermittency analysis has been hindered by the fact that factorial moments measurements corresponding to different scales are correlated, since the same data are conventionally used to calculate them. This invalidates many assumptions involved in fitting data sets and determining the best fit values of power-law exponents.
We present a novel approach to intermittency analysis, employing the well-established statistical and data science tool of Principal Component Analysis (PCA). This technique allows for the proper handling of correlations between scales without the need for subdividing the data sets available.
\end{abstract}




\section{Introduction}
\label{introduction}

A characteristic feature of a second order phase transition (expected to occur at the CP) is the divergence of the correlation length, leading to a scale-invariant system effectively described by a universality class. Of particular interest are local fluctuations of the order parameter of the QCD chiral phase transition; 
for an ideal case of an infinite size critical system, we expect fluctuations of the order parameter to be self-similar~\cite{vicsek1989fractal}, obeying power-laws with critical exponents determined by the 3D Ising universality class~\cite{Antoniou:2001,Antoniou:2005,Antoniou:2006}.

Such fluctuations correspond to a power-law scaling of the order parameter correlation function. In momentum space, the density-density correlations take the form~\cite{Antoniou:2016}:

\begin{equation}
    \langle \rho(\mathbf{k})\, \rho(\mathbf{k'}) \rangle \sim \vert \mathbf{k} - \mathbf{k'} \vert^{-f_2}
    \label{eq:corr_function}
\end{equation}
where $\rho(\mathbf{k})$ is the order parameter density in momentum space, $\mathbf{k - k'}$ is the momentum transfer, and $f_2$ is a power-law exponent determined by the universality class. 

Such correlations can be probed in transverse momentum space within the framework of an intermittency analysis of scaled factorial moments (SFMs), pioneered by Bia\l{}as, Peschanski and others~\cite{Bialas:1986,Bialas:1988,Bialas:1990xd,Antoniou:2006} as a method to detect non-trivial dynamical fluctuations in high energy nuclear collisions. In this analysis, we focus on the Second Scaled Factorial Moments (SSFM) $F_2(M)$. The method consists of partitioning an analysis window in D-dimensional momentum space into a number  of equal size bins (Fig.\ref{fig:gridcircles}~\emph{Left}), then examining how $F_2(M)$ of particle momenta scale with the total number $M^D$ of bins:

 \begin{equation}
 F_2(M) \equiv \left\langle \frac{1}{M^D} \displaystyle\sum_{i=1}^{M^D} n_i (n_i-1) \right\rangle \bigg/ \left\langle\frac{1}{M^D} \displaystyle\sum_{i=1}^{M^D} n_i \right\rangle^2
 \label{eq:F2M}
 \end{equation}
where $M$ is the number of bins per dimension, $n_i$ is the number of particles in the $i$-th bin, and $\langle\ldots\rangle$ denotes average over events. Commonly, D=2, i.e. we work in transverse momentum space. For a pure critical system, $F_2(M)$ is predicted to scale according to a power-law~\cite{Antoniou:2001,Antoniou:2005,Antoniou:2006,Antoniou:2016}:
\begin{equation}
 F_2(M) \sim \left(M^D\right)^{\varphi_2} \quad,\quad \varphi_2 \equiv \frac{f_2}{D}
 \label{eq:phi2_general}
 \end{equation}
 where the exponent $\varphi_2$ is called the intermittency index, and is connected to the density-density correlation function scaling exponent $f_2$.

In the conventional intermittency analysis approach (see e.g. Refs~\cite{NA49intermittency:2015, Davis:45ZFP, STAR:intermittency}), $F_2(M)$ are first calculated from an experimental data set; then, the moments can be fitted with a power-law function in the $M \gg 1$ region, to determine the best fit value of $\varphi_2$. However, this naive approach is problematic due to the fact that factorial moments measurements corresponding to different scales, i.e. different $M$-values, can be shown to be correlated, since the same data are conventionally used to calculate them. This invalidates the common assumption used in a simple $\chi^2$ fit, namely that errors per $M$-point are independent and identically distributed (i.i.d.). As a result, fit values and confidence intervals for $\varphi_2$ obtained under these assumptions are not reliable estimators of the true values.

A number of solutions to the problem of correlated bins have been proposed. As early as the 90s, the use of correlated $\chi^2$ fits for obtaining $\varphi_2$ has been proposed~\cite{Wosiek:1990}, using the full $M$-correlation matrix estimated e.g. via the bootstrap~\cite{Metzger,Efron:1979}; however, these fits are now known to be unstable, and to produce counter-intuitive results~\cite{Wosiek:1990,Michael:1994}. Another possibility is to use different subsets of events to calculate $F_2(M)$ for different $M$, eliminating bin correlations (see \cite{NA61SHINE:Tobiasz_intermittency} for an application of this technique); this has the drawback of severely reducing available event statistics per $M$-bin, and therefore inflating per-bin uncertainties considerably.

In the current work, we propose a different solution that employs the well-established statistical tool of \emph{Principal Component Analysis} (PCA)~\cite{Jolliffe2002}. PCA works by taking a set of (in general) correlated multi-dimensional observations, and determining the eigensystem of its covariance matrix. The matrix can then be diagonalized, and new variables, the Principal Components (PCs), can be defined in terms of the original ones (in our case, the $F_2(M)$), which are guaranteed to be statistically uncorrelated: the $\chi^2$ distribution of residuals (errors) in the PCs can be shown to follow the standard $\chi^2$ distribution with the corresponding number of degrees of freedom (ndf). A best fit to models can then be sought in this transformed PCs space, where simple $\chi^2$ can be defined with the assumption of point independence\footnote{Strictly speaking, the PC rotation being a linear transform can only eliminate \emph{linear} correlations; nevertheless, agreement of the empirical $\chi^2$ distribution of residuals with the standard $\chi^2$ distribution is sufficient for the purposes of simple $\chi^2$ fitting; we further discuss this point in Section~\ref{sec:PCA} and Appendix~\ref{ap:chi2_independence}}. The PCA methodology is closely related to, and a robust improvement of, the aforementioned correlated $\chi^2$ fit methodology; it allows one to control the influence of factorial moment sample noise on the fitting procedure, which is the main source of uncertainties in correlated $\chi^2$-fit. A detailed study of the connection between the two is provided in the Supplemental Material.

The layout of the present paper is as follows: in Section~\ref{sec:Methodology}, we describe the methodology used in the calculation of SSFMs and their uncertainties. Section~\ref{sec:CMC} gives a brief overview of the Critical Monte-Carlo (CMC) event generator~\cite{Antoniou:2001, Antoniou:2006, AMIAS_paper} suited for simulating power-law correlated particles in transverse momentum space, and used in this paper for the purpose of illustrating intermittency analysis methodology. Section~\ref{sec:PCA} presents and illustrates the use of the Principal Component Analysis (PCA) methodology, our proposed approach to solving the problem of $M$-bin correlations in intermittency analysis. Finally, in Section~\ref{sec:Summary} we present our summary and conclusions.

\section{Methodology}
\label{sec:Methodology}

 \begin{figure}
 \centering\includegraphics[width=.95\textwidth]{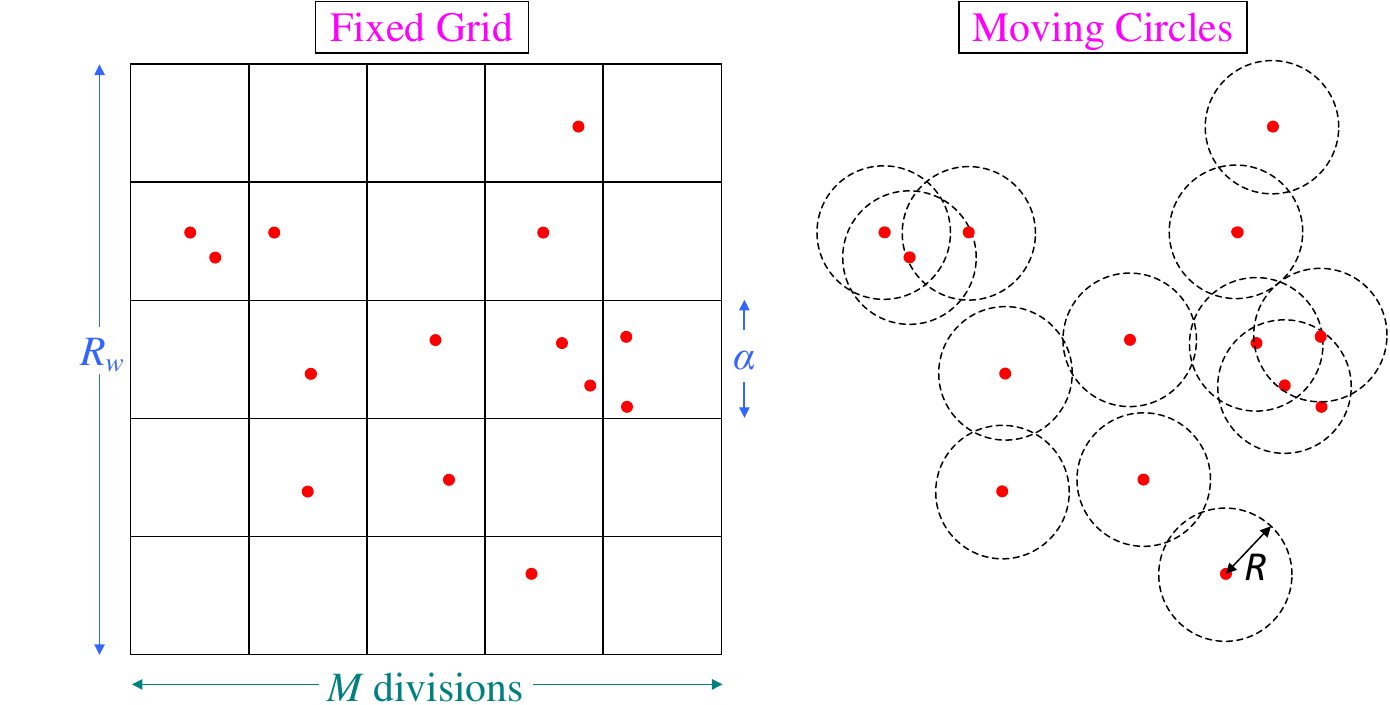}%
 \caption{\emph{Left:} Counting particle pairs in a transverse momentum space partitioning of $M \times M$ equal size bins. \emph{Right:} By centering circles around each point in the set, it is possible to account for all pairs of points within a given distance R; demanding equal area for bins and circles allows for a correspondence of scales: $\pi R^2 = \alpha^2$. Image reproduced from \cite{CorrelationIntegral2022}}
 \label{fig:gridcircles}
\end{figure}

We will describe intermittency methodology as applied in the proton channel, following published analyses by the NA49~\cite{NA49intermittency:2015} and NA61/SHINE Collaborations~\cite{Davis:45ZFP, NA61SHINE:Tobiasz_intermittency}. Analyses in other channels, such as dipions~\cite{NA49_pion_intermittency}, or even in all-charged~\cite{STAR:intermittency}, can equally benefit from the PCA methodology advanced in this paper. We choose to study the low event statistics case of SPS datasets as they better illustrate the benefits of the PCA methodology; results, if anything, would be improved with a high-statistics dataset, as all involved uncertainties would be reduced.

Proton intermittency analysis is performed on proton transverse momenta, with the purpose of probing the density-density correlation function, eq.(\ref{eq:corr_function}), of the net baryon number density $n_B$, of which the proton number density is a proxy~\cite{Hatta-Stephanov:2003}. For the idealized case of an infinite size system belonging to the 3D-Ising universality class, we predict~\cite{Antoniou:2006}:

\begin{equation}
    \langle n_B(\mathbf{k})\, n_B(\mathbf{k'}) \rangle \sim \vert \mathbf{k} - \mathbf{k'} \vert^{-5/3}
    \label{eq:baryon_corr}
\end{equation}
where $\mathbf{k - k'}$ is the momentum transfer.

Then, we can calculate the SSFMs, $F_2(M)$, eq.(\ref{eq:F2M}), in transverse momentum space ($D=2$). For a pure critical system belonging to the 3D-Ising universality class, $F_2(M)$ is predicted to follow a power-law~\cite{Antoniou:2006}:

 \begin{equation}
 F_2(M) \sim M^{2\cdot\varphi_{2,cr}} \quad,\quad \varphi_{2,cr}^{(p)} = 5/6
 \label{eq:phi2}
 \end{equation}
 where the exponent $\varphi_{2,cr}$ is the intermittency index defined in eq.~(\ref{eq:phi2_general}), and $\varphi_{2,cr}^{(p)}$ its theoretically predicted value for protons.

 In practice we study finite-size noisy systems, where many protons, in fact the vast majority, are not critically correlated. In that case, mixed event moments must be subtracted from the data moments in order to recover the critical component~\cite{NA49intermittency:2015}. This is a non-trivial operation; Critical Monte Carlo simulations~\cite{Antoniou:2001,Antoniou:2006} show that, under the experimentally common assumption of a dominant background, a correlator $\Delta F_2(M)$ can be defined~\cite{NA49intermittency:2015} that captures the critical contribution of the data:

  \begin{equation}
  \Delta F_2(M) \equiv F_2^{(d)}(M) - F_2^{(m)}(M)
 \label{eq:DF2}
 \end{equation}
where $F_2^{(d)}(M)$ are the original data moments and $F_2^{(m)}(M)$ is the approximation of background by mixed event moments.

Intermittent behavior, if present, will then be revealed in $\Delta F_2(M)$,
 \begin{equation}
  \Delta F_2 (M) \sim \left(M^2\right)^{\varphi_2},\, M \gg 1
  \label{eq:DF2_fit}
 \end{equation}
 Due to noise, finite-size effects, as well as the approximate nature of eq.~(\ref{eq:DF2}), the measured intermittency index $\varphi_2$ will in general differ from the predicted, eq.(\ref{eq:phi2}). Any power-law will only hold between a lower and upper momentum scale (equivalently, a maximum/minimum number of $M$-divisions). Experimental momentum resolution sets an absolute limit for the lower scale, of the order of $\sim 5$~MeV/$c$; the upper scale is imposed by the fact that eq.~(\ref{eq:baryon_corr}) only holds for small momentum transfer. For a typical analysis window of total width $\Delta p_{x,y} \simeq 3$~GeV/$c$, a power-law is sought in the range $32^2 \leq M^2 \leq 150^2$, equivalently $95 \geq \delta p_{x,y} \geq 20$~MeV/$c$.

 The usual methodology of computing $F_2(M)$ on a grid is computationally intensive, and introduces artifacts due to pair splitting by grid lines, when proton multiplicity per event is low. In the past, this problem was corrected for by introducing a lattice average over slightly displaced grids \cite{NA49intermittency:2015}; however, a computationally faster alternative to the lattice average has been developed \cite{CorrelationIntegral2022} using the correlation integral $C(R)$, defined as:

\begin{equation}
 C(R) = \frac{2}{\langle N_{mul} \left(N_{mul}-1\right)\rangle_{ev}} \Bigg\langle  \sum_{\substack{i,j \\ i < j}} \Theta\left( |x_i - x_j| \leq R \right)   \Bigg\rangle_{ev}
 \label{eq:corritg}
\end{equation}
where $R$ is a given length scale, $N_{mul}$ is the event multiplicity, and $\Theta$ is the step function, counting the number of pairs of particles in an event within a distance $R$ from each other. $C(R)$ can be calculated by (notionally) placing circles of radius $R$ around each of the points in the set (Fig.~\ref{fig:gridcircles}~\emph{Right}), and counting the points falling within them. We can then match the circle radius $R$ to the bin side length $\alpha$, and thus the number of divisions $M$ per dimension, by demanding equal area for bins and circles: $\pi R_M^2 = \alpha^2$. Thus, we arrive at a correspondence between $F_2(M)$ and $C(R_M)$,

\begin{equation}
 F_2(M) = \frac{\left\langle N_{mul} \left(N_{mul}-1 \right)\right\rangle_{ev}}{\left\langle N_{mul}\right\rangle_{ev}^2}M^2 C(R_M)
 \label{eq:corritg_F2_CR}
\end{equation}
that we can use to compute $F_2(M)$ efficiently, at the same time eliminating grid artifacts. The correlation integral is used throughout this paper for all experimental and Monte Carlo simulated data intermittency analyses.

SSFMs statistical errors are estimated via the bootstrap method~\cite{Metzger,Efron:1979}, which is a well-established statistical technique for obtaining unbiased error estimates of statistical quantities. It consists in randomly resampling the original set of events with replacement, hence constructing new sets equal in size to the original, where each original event can be drawn multiple times. By repeating this process, a large number of bootstrap samples ($\gtrsim 1000$) are created; then, $F_2(M)$ are calculated for each bootstrap sample in the same manner as for the original. The resulting values can be used to obtain the bootstrap $F_2(M)$ statistical distribution, as well as its standard error, confidence intervals, or any other measure of variance desirable. It also lends itself naturally and effortlessly to calculating the correlation matrix of $F_2(M)$ between different bins $M$.
 
 Bootstrap is straightforward to implement, only requiring calculation of the original statistics (the SSFMs), in contrast to error propagation, which requires calculating higher moments~\cite{Metzger}. Simulations show that bootstrap estimates of variance and covariance of SSFMs can be trusted; however, the centroids (average, median) estimated by bootstrap will certainly be biased towards the original sample. Thus, a simple least-squares fit of bootstrap-averaged $F_2 (M)$ values is still invalid. We will address this issue, as well as methods to deal with it, in Section~\ref{sec:PCA}. 

\section{The Critical Monte Carlo simulation}
\label{sec:CMC}

In order to systematically explore the behaviour and scaling of SSFMs of noisy experimental datasets, we need a way of simulating critical correlations between protons, non-critical background protons, as well as their interplay. Monte Carlo simulations provide a royal road towards such insight. 

For this purpose, we use a modified version of the Critical Monte-Carlo
(CMC) event generator~\cite{Antoniou:2001, Antoniou:2006, AMIAS_paper} suited for simulating protons in transverse momentum space; simulated protons are produced by sampling a truncated L\'evy
walk process to exhibit density-density correlations mimicking those originating from a fireball freezing out at the QCD critical point\footnote{Here, we refer to the chemical freeze-out of the system, when critical proton correlations are established along the hadronization process.}. The power-law
exponent is adjustable within a range of values; for example, it can be chosen to describe correlations characterizing a critical system in
the 3d-Ising universality class. L\'evy walk step bounds can be fine-tuned in order to produce power-law correlations within the desired momentum scales. The associated intermittency index range currently achievable is $\phi_2 \in [0.1, 1]$; the value $\phi_2 = 5/6$ corresponds to a fractal mass dimension of $d_F = 1/3$ for the 2-dimensional L\'evy walk. Random walk cluster centers are drawn from an empirical 2-dimensional average per event proton $p_{x,y}$ distribution. This, as well as the per-event proton multiplicity distribution, can be directly plugged in the Monte Carlo.

A number of uncorrelated proton momenta drawn from a plugged-in one-particle $p_T$
distribution replace the critical protons with an adjustable probability per particle. These simulate the effect of non-critical background contamination
on the critical signal, with the desired background level.  It must be emphasized that CMC is a purely static simulation with no aspirations to describe the dynamic evolution of the colliding nuclei; only momenta correlations and the final distribution of momenta are simulated, through the use of critically correlated and background protons. Additionally, the generated CMC events can be further processed by including transformations of momenta simulating detector effects, for better comparison with experimental data.

\begin{figure}
 \centering\includegraphics[width=.5\textwidth]{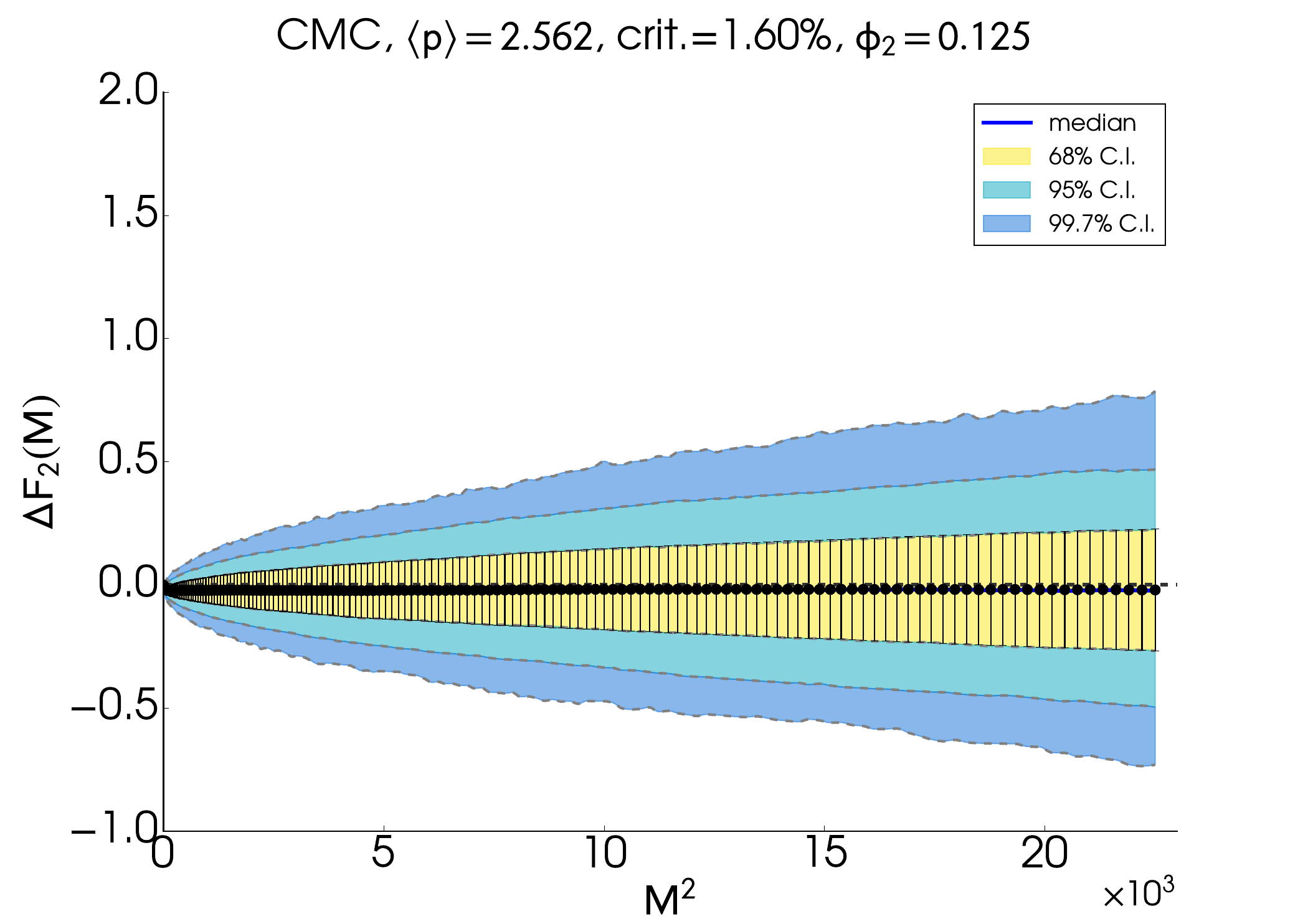}%
 \includegraphics[width=.5\textwidth]{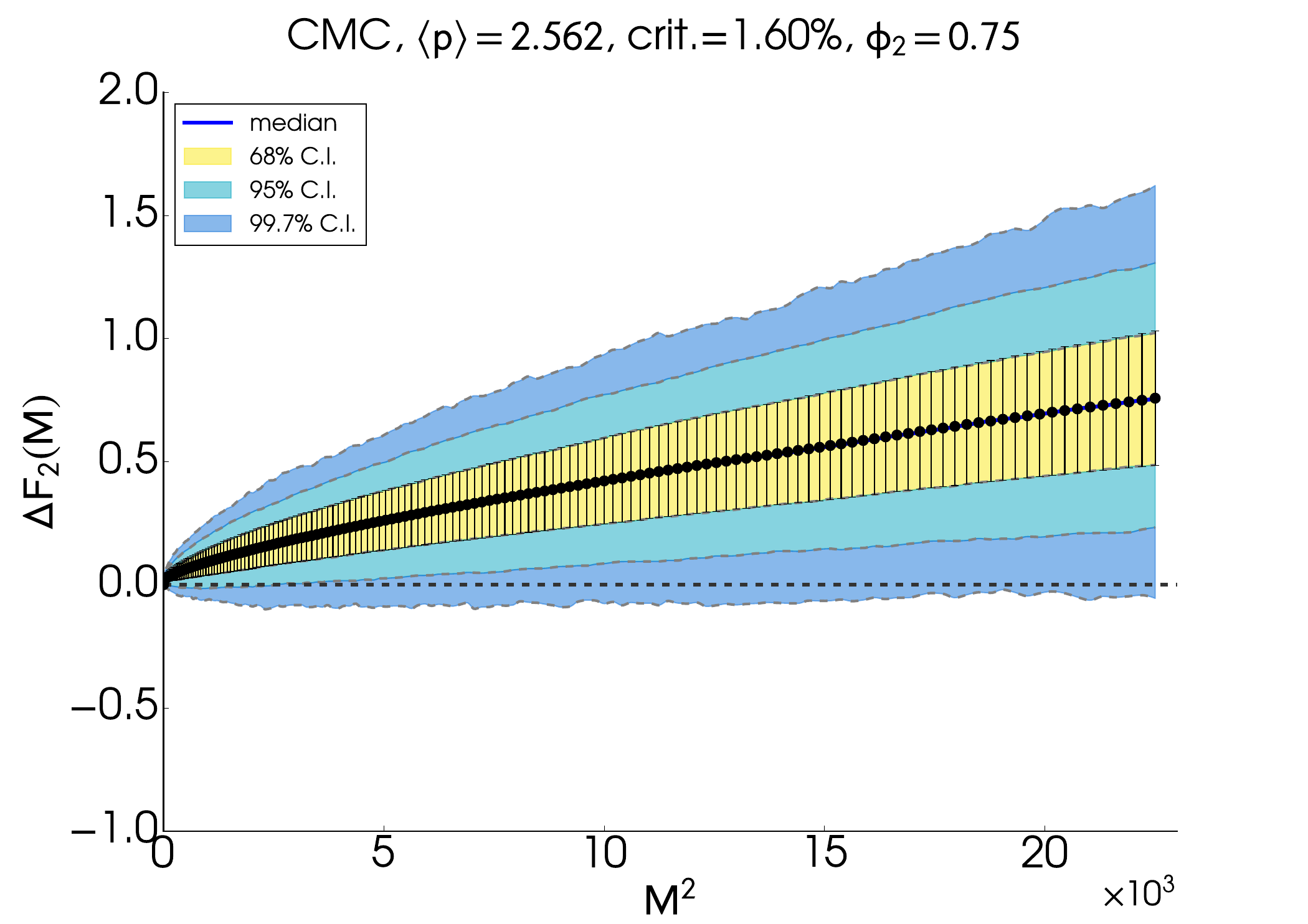}
 \caption{Correlator $\Delta F_2(M)$ for two sets of $\sim400\text{K}$ CMC-simulated non-central Ar+Sc collisions at SPS beam momentum 150$A$~GeV/$c$ ($\sqrt{s_{NN}} \approx 16.8$~GeV). For both sets, the percentage of critical to total simulated protons has been set to 1.60\%. Critical exponent (intermittency index) is set to $\phi_2 = 0.125$ \emph{(left)} and $\phi_2 = 0.75$ \emph{(right)} respectively. The black points correspond to the average $\Delta F_2(M)$ of $\sim8\text{K}$ independent iterations (samples) of the simulation. Colored bands correspond to 1-(\emph{yellow}), 2-(\emph{light blue}), and 3-$\sigma$ (\emph{dark blue}) confidence intervals of independent sample variation, respectively.}
 \label{fig:CMC_ArSc150_examples}
\end{figure}

As a working example, CMC can be tuned to simulate non-central Ar+Sc collisions at the top SPS energy ($\sqrt{s_{NN}} \approx 16.8$~GeV), using empirical proton $p_{x,y}$ and multiplicity distributions and simulated detector effects drawn from e.g. the NA61/SHINE experiment~\cite{NA61SHINE:Tobiasz_intermittency}. Fig.~\ref{fig:CMC_ArSc150_examples} shows two examples of the correlator $\Delta F_2(M)$ obtained by the CMC simulation with the aforementioned settings. Two different values of the intermittency index exponent $\phi_2$ have been used, one low \emph{(left)} and one close to the critical prediction \emph{(right)}. In each case, the factorial moments are calculated on sets of $\sim400\text{K}$ events, similar to NA61/SHINE available statistics, and simulation is repeated independently for $\sim8\text{K}$ iterations, keeping the same simulation parameters throughout. Independent simulations allow us to estimate the variability of the resulting $\Delta F_2(M)$ at this level of event statistics; the average $\Delta F_2(M)$ gives us the overall trend of the model. We note that the qualitative behavior of the correlator is similar to that of the conventional Ar+Sc NA61/SHINE analysis~\cite{Davis:45ZFP}. We also note a very similar internal spread of values as obtained for Ar+Sc NA61/SHINE via the bootstrap method, with the crucial difference that now the different samples are statistically independent.

\section{Principal Component Analysis}
\label{sec:PCA}

The factorial moments $\Delta F_2^\text{exp}(M)$ of an experimental or synthetic data set should be compared against the corresponding moments produced or predicted by a model, in order to obtain useful information about the scaling behaviour, and in particular the intermittency index $\varphi_2$ value range, that is compatible with the behaviour exhibited by the experimental data set in question. This may be done in a variety of ways; the most straightforward one is to directly fit experimental $\Delta F_2(M)$ with a power-law function, e.g.:

\begin{equation}
\Delta F_2^{\text{power-law}}(M) \equiv 10^{\alpha_0} \left( \frac{M^2}{10^4} \right)^{\varphi_2},
    \label{eq:power_law_a0_phi2}
\end{equation}
where $\alpha_0$ and $\varphi_2$ are parameters quantifying the magnitude and power-law exponent of the fitted power-law function, respectively. The best fit parameter values are found by minimizing the square $\chi^2$ of the residuals:

\begin{equation}
\chi^2 = \sum_M \left( \frac{\Delta F_2^\text{power-law}(M) - \Delta F_2^\text{exp}(M)}{\sigma(M)} \right)^2
    \label{eq:chisquare}
\end{equation}
where $\sigma(M)$ is a measure of the $M$-bin moment uncertainty, estimated e.g. via the bootstrap. As mentioned in Section~\ref{introduction}, however, this approach gives biased results due to the false assumption of independent (uncorrelated) $M$-bins. Furthermore, we would often like to obtain more than the maximum likelihood values of the model parameters -- we would like to obtain their whole joint probability distribution, based on the experimental data set.

This can be achieved by calculating $\chi^2$ as a function of fit parameters; subsequently, its weight $\exp(-\chi^2/2)$ can be normalized by integrating over all considered parameter values\footnote{This parameter weighting method is largely inspired by the more general AMIAS scheme; for a detailed analysis, see Refs~\cite{Stiliaris:2007, Papanicolas:2012}.}. However, $\chi^2$ cannot be calculated in the original $M$-bins, due to their correlations. The solution, provided by Principal Component Analysis (PCA)~\cite{Jolliffe2002}, is to use knowledge of the covariance matrix of $\Delta F_2(M)$, $cov\left( F_2(M) \;,\; F_2(N)\right)$ in order to diagonalize it, and define linear combinations of $\Delta F_2(M)$ among different $M$ such that $cov\left( F_2(PC_M) \;,\; F_2(PC_N)\right)$ is diagonal;  where $PC_M, PC_N$ correspond to all possible pairs of Principal Components (PCs) defined by the original $M$-bins. This process can be viewed as a geometric ``rotation'' of an $\#M$-dimensional cloud of points $\Delta F_2(M)$ into a new orientation of coordinates, corresponding to the PCs (the ``eigenvectors'' of the covariance matrix), which progressively account for more and more of the variance present in the dataset; i.e., the PCs are arranged in order of decreasing eigenvalues.

\begin{figure}[htb]
 \centering%
 \includegraphics[width=0.425\textwidth]{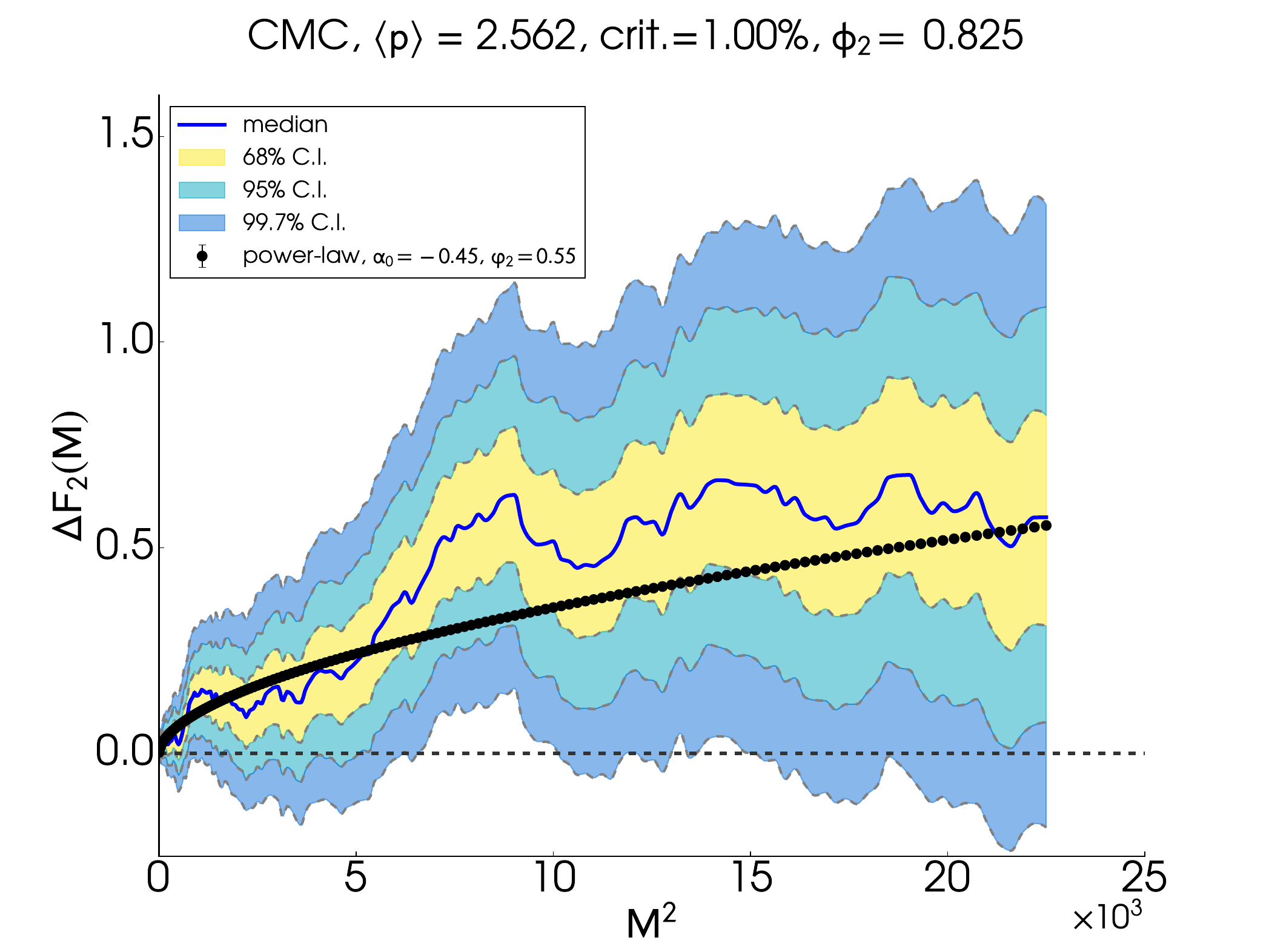}%
 \includegraphics[width=0.275\textwidth]{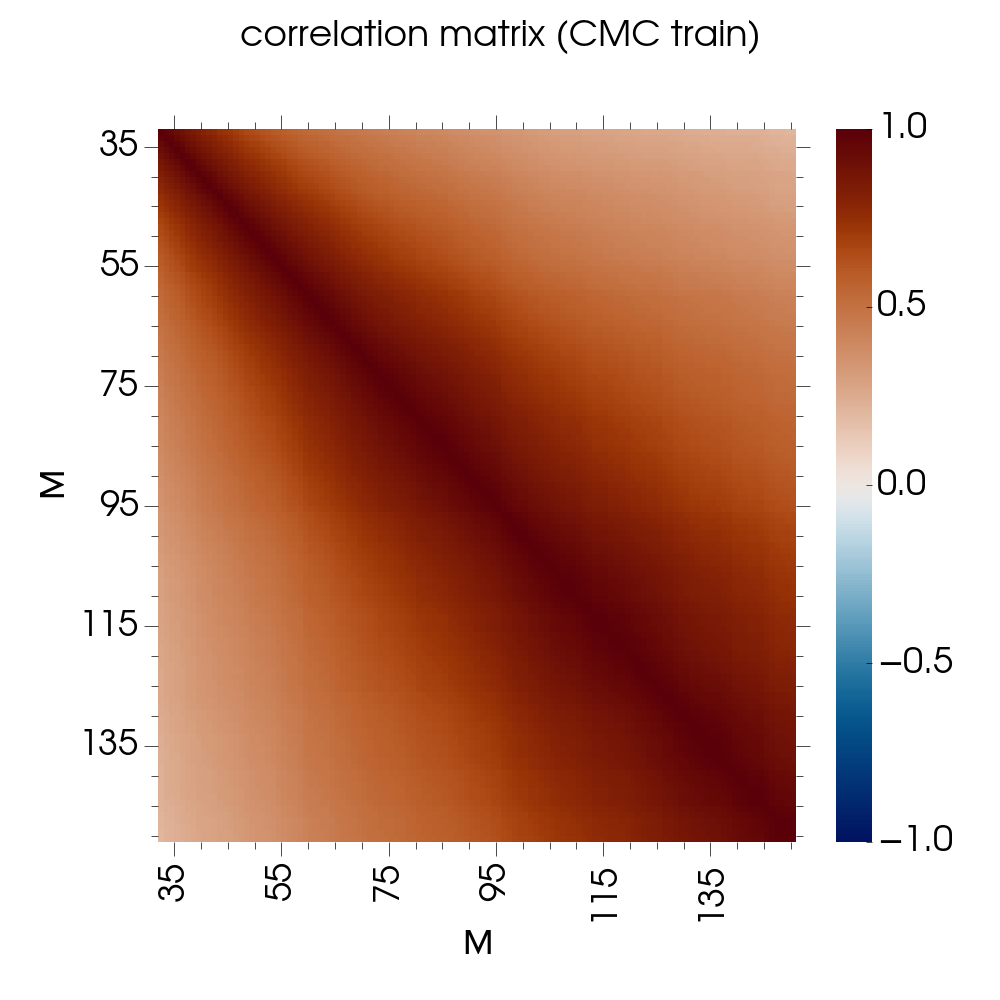}%
 \includegraphics[width=0.275\textwidth]{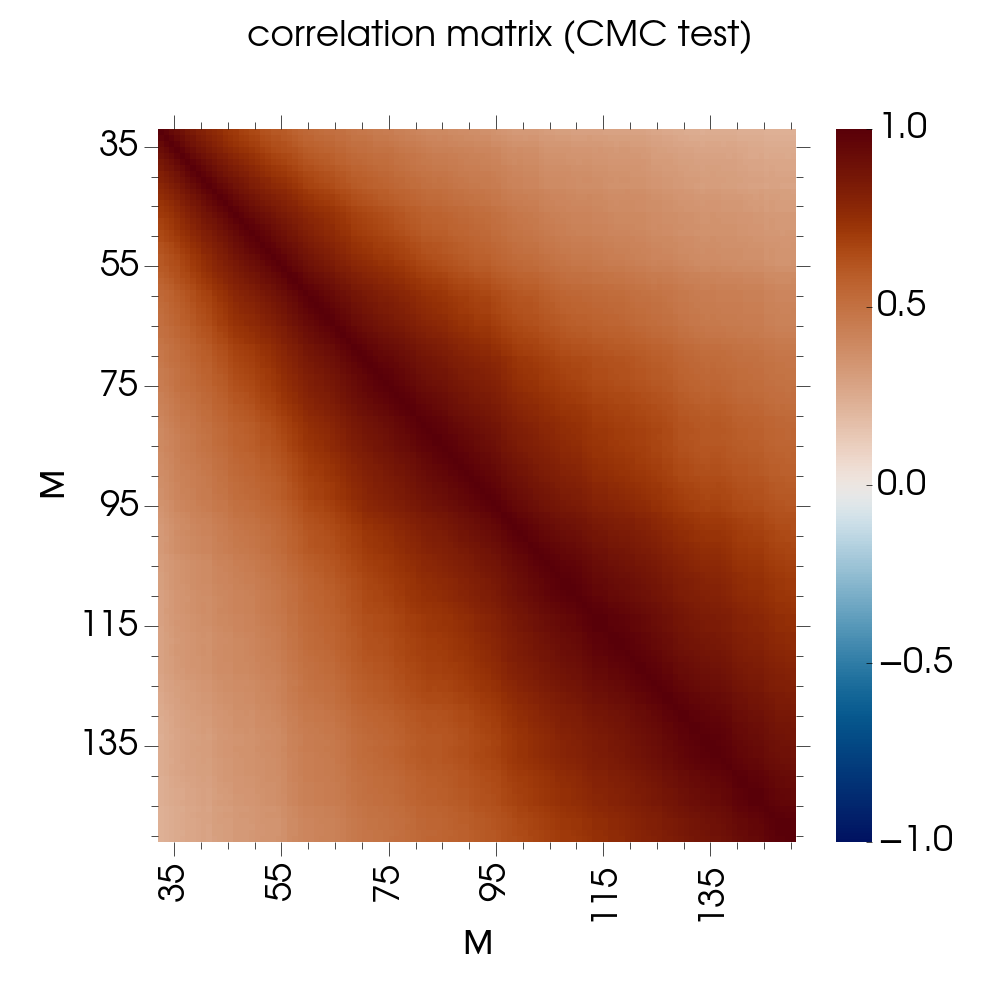}\\%
 \centering%
 \includegraphics[width=0.425\textwidth]{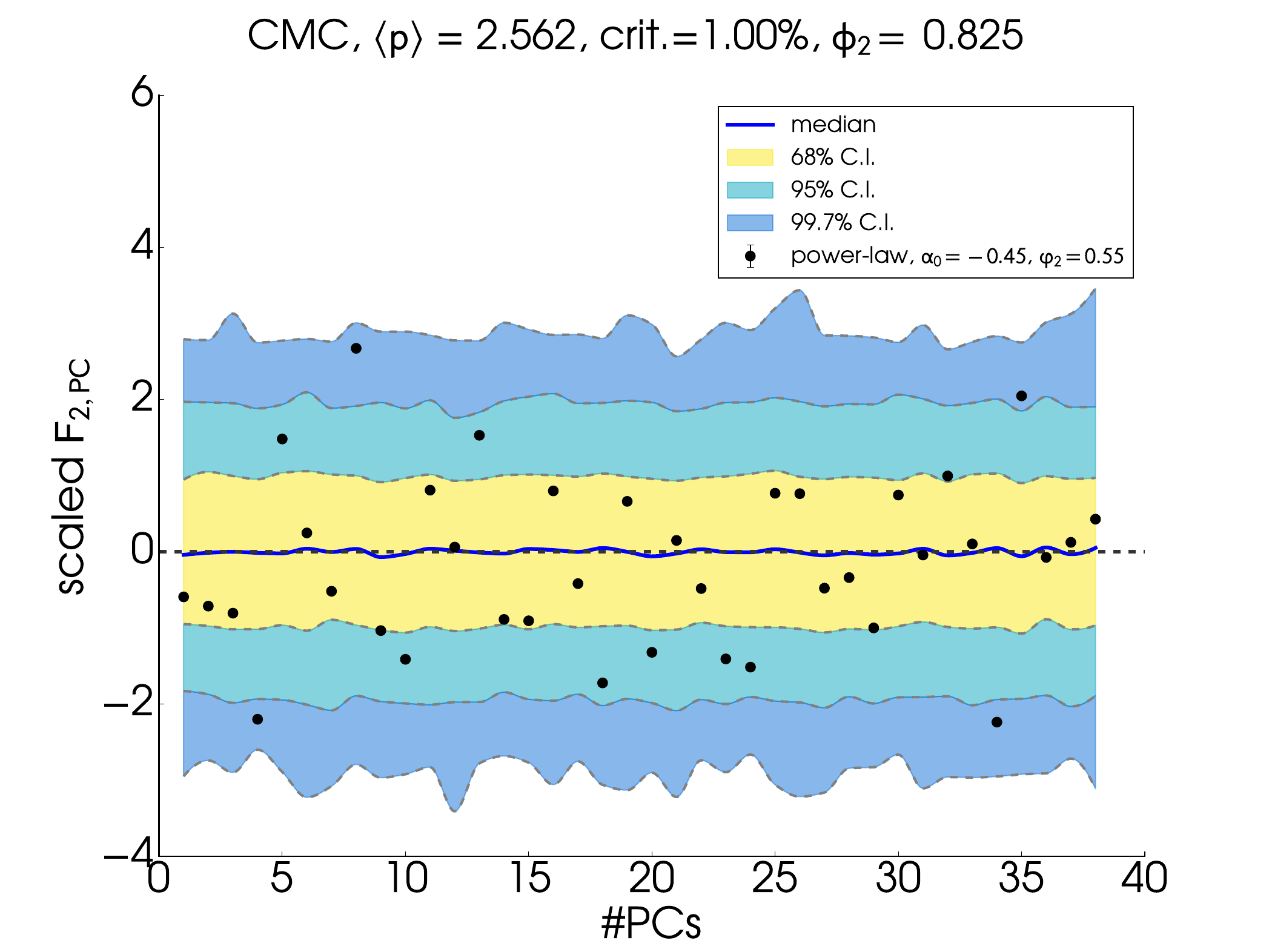}%
 \includegraphics[width=0.275\textwidth]{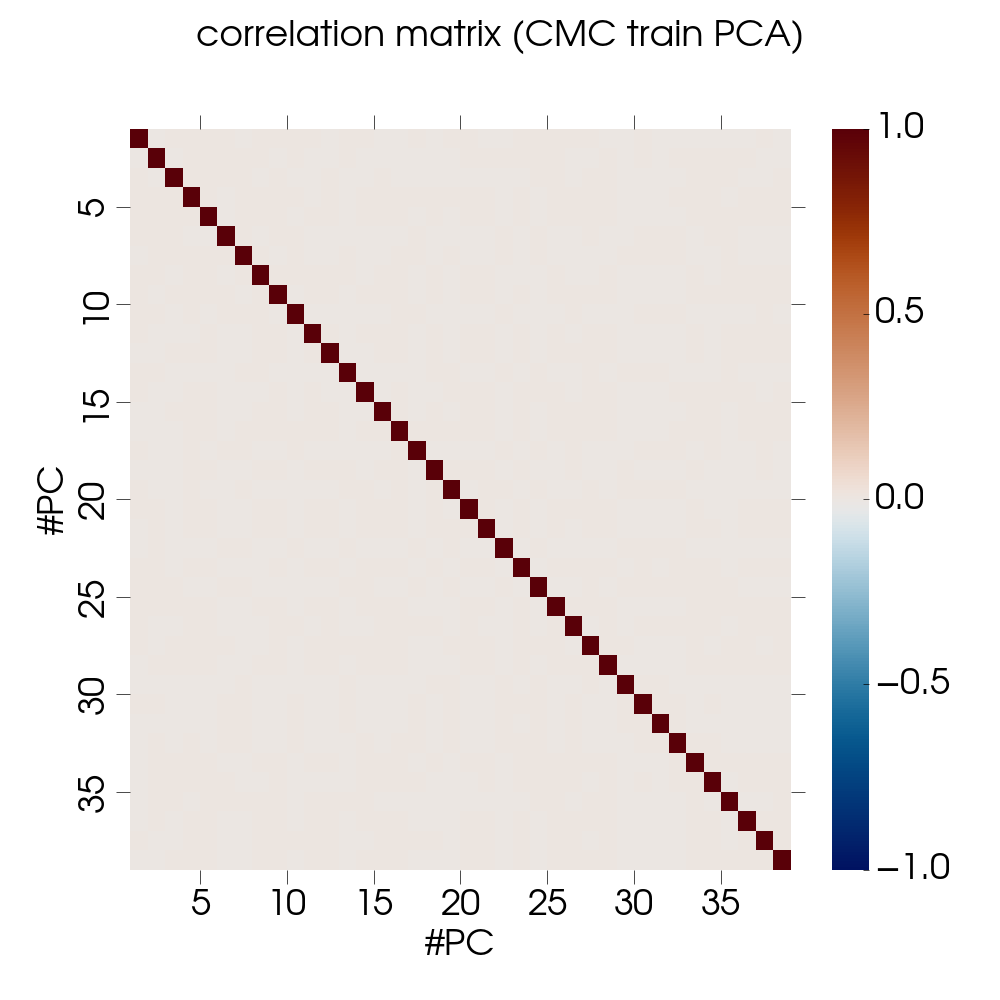}%
 \includegraphics[width=0.275\textwidth]{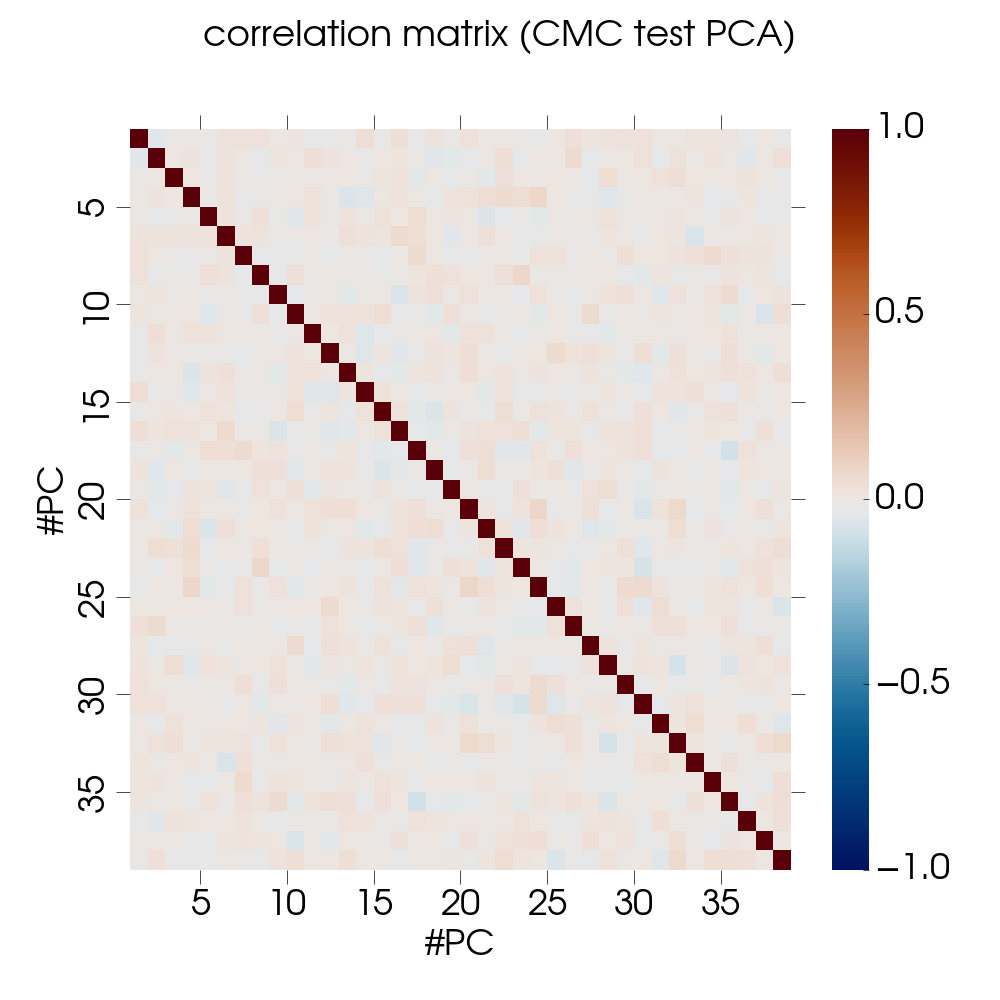}
 \caption{\emph{(Top Row)} \emph{Left:} Correlator $\Delta F_2(M)$ of $\sim8\text{K}$ bootstrap samples of a synthetic CMC data set (colored bands) corresponding to $\sim400\text{K}$ Ar+Sc-like collision events containing proton particles. A power-law model parametrized in $(\alpha_0,\varphi_2)$, eq.~(\ref{eq:power_law_a0_phi2}), is also plotted (black points); \emph{Middle/Right:} The correlation matrix between $M$-bins, for the train and test portions of the CMC dataset, respectively. \emph{(Bottom Row)} \emph{Left:} The correlator $\Delta F_2(M)$ for the same CMC data set, transformed to the PC coordinates (38 most significant components kept); \emph{Middle/Right:} The correlation matrix between PCs, for the train and test portions of the CMC dataset, respectively. Colored bands correspond to 1-(\emph{yellow}), 2-(\emph{light blue}), and 3-$\sigma$ (\emph{dark blue}) confidence intervals of bootstrap sample variation, respectively. Blue median line is the original dataset $\Delta F_2(M)$.}
 \label{fig:PCA_application}
\end{figure}

Fig.~\ref{fig:PCA_application} illustrates how PCA is applied in practice. On the top left panel, $\Delta F_2(M)$ of a synthetic Critical Monte Carlo dataset of $\sim400\text{K}$ Ar+Sc-like collision events (dark blue line), containing proton particles, are plotted in the original $M$-bin coordinates, along with the distribution of $\sim8\text{K}$ of their bootstrap samples (colored bands). First, the bootstrap samples are randomly partitioned  into ``train''/``test'' subsets, at 80\% / 20\%  rates, respectively. Then, we plot the correlation matrices between $M$-bins for both subsets (middle and right of top row). We observe considerable variation of bootstrap samples vs the original; moreover, these variations are strongly correlated across neigbouring $M$-bins, as shown by the correlation matrices. Along the CMC, the $\Delta F_2(M)$ of a particular power-law model, eq.~(\ref{eq:power_law_a0_phi2}), are plotted (black points); these are to be compared against the CMC original moments (dark blue line).

We use the ``train'' bootstrap subset to determine the PCs, i.e. linear combinations of original $M$-bins, that diagonalize the correlation matrix. If we then rotate and scale both the CMC original \& bootstrap samples, and the power-law model points, into the 38 most significant PCs\footnote{See Appendix~\ref{ap:optimal_PCs} for how the optimal number of PCs is determined.}, we obtain the bottom-left plot, Fig.~\ref{fig:PCA_application}. The original CMC sample now corresponds to the zero line, whereas the bootstrap variations have transformed to 1-2-3-$\sigma$ horizontal bands. PC bins are independently varying, as is illustrated by the ``train''/``test'' correlation matrices in PC-space (middle/right of bottom row, respectively). It is worth noting that the transformation, determined by the train set, almost perfectly diagonalizes the test set, confirming the stability of PCA rotation across different bootstrap subsets. The $\chi^2$ of power-law model (black points) can now be calculated as the simple sum of squares of $F_{2,\text{PC}}$ of the power-law points across the PCs, which is unbiased due to the independence of different PC variations.

An important caveat is in order here. The statistical independence of PCs is limited to linear correlations, which are the only ones detected by the covariance/correlation matrix, and eliminated by the PCA rotation. It is therefore crucial to the validity of the PCA methodology to check that higher order correlations, if present, should be negligible. This assumption can be tested in a straightforward manner, by comparing the empirical distribution of bootstrap or Monte Carlo sample residuals (colored bands in Fig.~\ref{fig:PCA_application} top/bottom left) to a standard $\chi^2$ distribution with the appropriate number of degrees of freedom (equal to the number of summed components, eq.~(\ref{eq:chisquare})). An empirical $\chi^2$ distribution that closely matches the standard $\chi^2$ distribution to a good approximation justifies the assumption of PC independence. See Appendix~\ref{ap:chi2_independence} for a detailed description of the test, as well as its application to the synthetic CMC data set, Fig.~\ref{fig:PCA_application}.

It must be noted that the roles of experimental and model $\Delta F_2(M)$ can be interchanged for the purposes of eq.~(\ref{eq:chisquare}); when we have access to independent variations of the model's $\Delta F_2(M)$, as with CMC, we can determine the PCA rotation based on the model's correlation matrix, then transform the experimental (original) dataset to the same coordinates; in this case, the zero baseline corresponds to the average model. In fact, this option is preferable, since model variations are independent, whereas bootstrap ones are not; in both cases, the baseline $\sigma(M)$ for uncertainties, eq.~(\ref{eq:chisquare}), should correspond to the \emph{largest} uncertainties between model \& experimental data, or an aggregate uncertainty derived from both.

\begin{figure}[htb]
    \centering%
 \includegraphics[width=.8\textwidth]{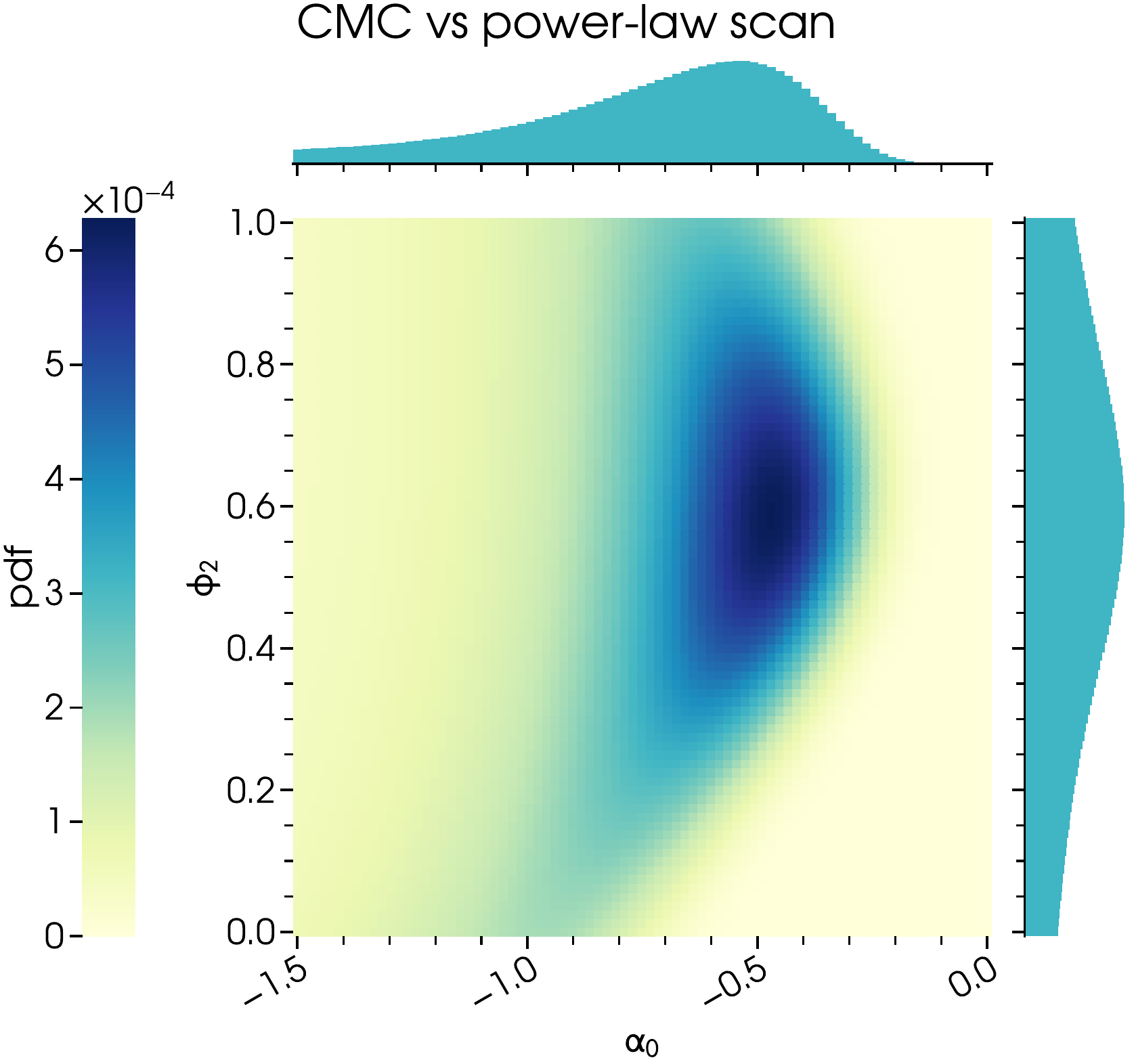}%
    \caption{Normalized likelihood $\exp(-\chi^2/2) \,\bigg/\, \sum \exp(-\chi^2/2)$ for a grid of power-law parameters vs synthetic CMC dataset plotted in Fig.~\ref{fig:PCA_application}. Marginal $\alpha_0$ ($\varphi_2$) distribution plotted on the top (right) of 2D-plot.}
    \label{fig:CMC_plaw_distribution}
\end{figure}

Fig.~\ref{fig:CMC_plaw_distribution} shows a map of the normalized likelihoods of a collection of power-law models, parametrized in $(\alpha_0, \varphi_2)$-space, vs the synthetic CMC dataset plotted in Fig.~\ref{fig:PCA_application}. Individual $\chi^2$ values are obtained in PC-space as illustrated in Fig.~\ref{fig:PCA_application}; normalization is performed by dividing by the sum of all model weights. We observe a unimodal 2D-distribution, with a maximum located at the best fit model\footnote{Due to the presence of background protons, the best fit value for $\varphi_2$ differs from the plugged-in value of $\varphi_2 = 0.825$.}. The width of the distribution can be used to obtain confidence intervals for the most likely power-law models, given the data.

\section{Summary and conclusions}
\label{sec:Summary}

Intermittency analysis of factorial moments is a promising observable for the detection of the critical point of strongly interacting matter, provided that it is used in the appropriate manner. When trying to determine which parameters, in particular which power-laws, are compatible with a given data set, correlations between different momentum scales cannot be ignored; they can be shown to be significant and long-range. We have adapted the well-established statistical and data science tool of Principal Component Analysis (PCA) to intermittency studies, and demonstrated it capable of both decoupling correlation between scales and reducing the effective dimensionality of factorial moment datasets. This is achieved without decimating the available statistics (number of collision events). Coupling PCA with a scan of models in parameter space, reliable likelihoods can be obtained for each individual model; moreover, the joint likelihood distribution of all model parameters can be extracted, and unbiased confidence intervals for each parameter drawn, allowing for better interpretation of the significance of experimental results, and the approach of the corresponding system to the critical point.

\section*{Acknowledgements}
The author would like to thank prof F.~K.~Diakonos for the initial encouragement to release this analysis, as well as for providing essential feedback. I am grateful to prof A.~K.~Rybicki and dr A.~Marcinek, as well as prof E.~Stiliaris and dr A.~S.~Kapoyannis, for many fruitful discussions invaluable to the development of the ideas presented in this paper.

%% Bibliography
\bibliographystyle{elsarticle-num}
\bibliography{my-bib-database}

\begin{thebibliography}{10}
\expandafter\ifx\csname url\endcsname\relax
  \def\url#1{\texttt{#1}}\fi
\expandafter\ifx\csname urlprefix\endcsname\relax\def\urlprefix{URL }\fi
\expandafter\ifx\csname href\endcsname\relax
  \def\href#1#2{#2} \def\path#1{#1}\fi

\bibitem{vicsek1989fractal}
T.~Vicsek, \href{https://books.google.pl/books?id=N8XSq7rTzTkC}{Fractal Growth
  Phenomena}, World Scientific, 1989.
\newline\urlprefix\url{https://books.google.pl/books?id=N8XSq7rTzTkC}

\bibitem{Antoniou:2001}
N.~G. Antoniou, Y.~F. Contoyiannis, F.~K. Diakonos, A.~I. Karanikas, C.~N.
  Ktorides, {Pion production from a critical QCD phase}, Nucl. Phys. A 693
  (2001) 799--824.
\newblock \href {http://arxiv.org/abs/hep-ph/0012164}
  {\path{arXiv:hep-ph/0012164}}, \href
  {https://doi.org/10.1016/S0375-9474(01)00921-6}
  {\path{doi:10.1016/S0375-9474(01)00921-6}}.

\bibitem{Antoniou:2005}
N.~G. Antoniou, Y.~F. Contoyiannis, F.~K. Diakonos, G.~Mavromanolakis,
  {Critical QCD in nuclear collisions}, Nucl. Phys. A 761 (2005) 149--161.
\newblock \href {http://arxiv.org/abs/hep-ph/0505185}
  {\path{arXiv:hep-ph/0505185}}, \href
  {https://doi.org/10.1016/j.nuclphysa.2005.07.003}
  {\path{doi:10.1016/j.nuclphysa.2005.07.003}}.

\bibitem{Antoniou:2006}
N.~G. Antoniou, F.~K. Diakonos, A.~S. Kapoyannis, K.~S. Kousouris, {Critical
  opalescence in baryonic QCD matter}, Phys. Rev. Lett. 97 (2006) 032002.
\newblock \href {http://arxiv.org/abs/hep-ph/0602051}
  {\path{arXiv:hep-ph/0602051}}, \href
  {https://doi.org/10.1103/PhysRevLett.97.032002}
  {\path{doi:10.1103/PhysRevLett.97.032002}}.

\bibitem{Antoniou:2016}
N.~G. Antoniou, N.~Davis, F.~K. Diakonos, {Fractality in momentum space: a
  signal of criticality in nuclear collisions}, Phys. Rev. C 93~(1) (2016)
  014908.
\newblock \href {http://arxiv.org/abs/1510.03120} {\path{arXiv:1510.03120}},
  \href {https://doi.org/10.1103/PhysRevC.93.014908}
  {\path{doi:10.1103/PhysRevC.93.014908}}.

\bibitem{Bialas:1986}
A.~Bialas, R.~B. Peschanski, {Moments of Rapidity Distributions as a Measure of
  Short Range Fluctuations in High-Energy Collisions}, Nucl. Phys. B 273 (1986)
  703--718.
\newblock \href {https://doi.org/10.1016/0550-3213(86)90386-X}
  {\path{doi:10.1016/0550-3213(86)90386-X}}.

\bibitem{Bialas:1988}
A.~Bialas, R.~B. Peschanski, {Intermittency in Multiparticle Production at
  High-Energy}, Nucl. Phys. B 308 (1988) 857--867.
\newblock \href {https://doi.org/10.1016/0550-3213(88)90131-9}
  {\path{doi:10.1016/0550-3213(88)90131-9}}.

\bibitem{Bialas:1990xd}
A.~Bialas, R.~C. Hwa, {Intermittency parameters as a possible signal for quark
  - gluon plasma formation}, Phys. Lett. B 253 (1991) 436--438.
\newblock \href {https://doi.org/10.1016/0370-2693(91)91747-J}
  {\path{doi:10.1016/0370-2693(91)91747-J}}.

\bibitem{NA49intermittency:2015}
T.~Anticic, et~al., {Critical fluctuations of the proton density in A+A
  collisions at 158$A$ GeV}, Eur. Phys. J. C 75~(12) (2015) 587.
\newblock \href {http://arxiv.org/abs/1208.5292} {\path{arXiv:1208.5292}},
  \href {https://doi.org/10.1140/epjc/s10052-015-3738-5}
  {\path{doi:10.1140/epjc/s10052-015-3738-5}}.

\bibitem{Davis:45ZFP}
N.~Davis, {Searching for the Critical Point of Strongly Interacting Matter in
  Nucleus\textendash{}Nucleus Collisions at CERN SPS}, Acta Phys. Polon. Supp.
  13~(4) (2020) 637--643.
\newblock \href {http://arxiv.org/abs/2002.06636} {\path{arXiv:2002.06636}},
  \href {https://doi.org/10.5506/APhysPolBSupp.13.637}
  {\path{doi:10.5506/APhysPolBSupp.13.637}}.

\bibitem{STAR:intermittency}
M.~Abdulhamid, et~al., {Energy dependence of intermittency for charged hadrons
  in Au+Au collisions at RHIC}, Phys. Lett. B 845 (2023) 138165.
\newblock \href {http://arxiv.org/abs/2301.11062} {\path{arXiv:2301.11062}},
  \href {https://doi.org/10.1016/j.physletb.2023.138165}
  {\path{doi:10.1016/j.physletb.2023.138165}}.

\bibitem{Wosiek:1990}
B.~Wosiek, {Intermittency analysis of correlated data}, Acta Phys. Polon. B 21
  (1990) 1021--1030.

\bibitem{Metzger}
W.~J. Metzger, Estimating the uncertainties of factorial moments, hEN-455
  (unpublished) (2004).

\bibitem{Efron:1979}
B.~Efron, \href{https://doi.org/10.1214/aos/1176344552}{{Bootstrap Methods:
  Another Look at the Jackknife}}, The Annals of Statistics 7~(1) (1979) 1 --
  26.
\newblock \href {https://doi.org/10.1214/aos/1176344552}
  {\path{doi:10.1214/aos/1176344552}}.
\newline\urlprefix\url{https://doi.org/10.1214/aos/1176344552}

\bibitem{Michael:1994}
C.~Michael, {Fitting correlated data}, Phys. Rev. D 49 (1994) 2616--2619.
\newblock \href {http://arxiv.org/abs/hep-lat/9310026}
  {\path{arXiv:hep-lat/9310026}}, \href
  {https://doi.org/10.1103/PhysRevD.49.2616}
  {\path{doi:10.1103/PhysRevD.49.2616}}.

\bibitem{NA61SHINE:Tobiasz_intermittency}
H.~Adhikary, et~al., {Search for the critical point of strongly-interacting
  matter in $^{40}$Ar ~+~$^{45}$Sc collisions at 150A ~Ge V /c using scaled
  factorial moments of protons}, Eur. Phys. J. C 83~(9) (2023) 881.
\newblock \href {http://arxiv.org/abs/2305.07557} {\path{arXiv:2305.07557}},
  \href {https://doi.org/10.1140/epjc/s10052-023-11942-9}
  {\path{doi:10.1140/epjc/s10052-023-11942-9}}.

\bibitem{Jolliffe2002}
I.~Jolliffe, \href{https://doi.org/10.1007/b98835}{Principal Component
  Analysis}, Springer New York, New York, NY, 2002.
\newblock \href {https://doi.org/10.1007/b98835} {\path{doi:10.1007/b98835}}.
\newline\urlprefix\url{https://doi.org/10.1007/b98835}

\bibitem{AMIAS_paper}
N.~G. Antoniou, et~al., {Decoding the QCD critical behaviour in A + A
  collisions}, Nucl. Phys. A 1003 (2020) 122018.
\newblock \href {http://arxiv.org/abs/2004.12133} {\path{arXiv:2004.12133}},
  \href {https://doi.org/10.1016/j.nuclphysa.2020.122018}
  {\path{doi:10.1016/j.nuclphysa.2020.122018}}.

\bibitem{CorrelationIntegral2022}
F.~K. Diakonos, A.~S. Kapoyannis, {Correlation integral vs. second order
  factorial moments and an efficient computational technique}, Eur. Phys. J. C
  82~(3) (2022) 200.
\newblock \href {http://arxiv.org/abs/2109.12571} {\path{arXiv:2109.12571}},
  \href {https://doi.org/10.1140/epjc/s10052-022-10098-2}
  {\path{doi:10.1140/epjc/s10052-022-10098-2}}.

\bibitem{NA49_pion_intermittency}
T.~Anticic, et~al.,
  \href{https://link.aps.org/doi/10.1103/PhysRevC.81.064907}{{Search for the
  QCD critical point in nuclear collisions at $158A$ GeV at the CERN Super
  Proton Synchrotron (SPS)}}, Phys. Rev. C 81 (2010) 064907.
\newblock \href {https://doi.org/10.1103/PhysRevC.81.064907}
  {\path{doi:10.1103/PhysRevC.81.064907}}.
\newline\urlprefix\url{https://link.aps.org/doi/10.1103/PhysRevC.81.064907}

\bibitem{Hatta-Stephanov:2003}
Y.~Hatta, M.~A. Stephanov, {Proton number fluctuation as a signal of the QCD
  critical endpoint}, Phys. Rev. Lett. 91 (2003) 102003, [Erratum:
  Phys.Rev.Lett. 91, 129901 (2003)].
\newblock \href {http://arxiv.org/abs/hep-ph/0302002}
  {\path{arXiv:hep-ph/0302002}}, \href
  {https://doi.org/10.1103/PhysRevLett.91.102003}
  {\path{doi:10.1103/PhysRevLett.91.102003}}.

\bibitem{Stiliaris:2007}
E.~Stiliaris, C.~N. Papanicolas, {Multipole extraction: A Novel, model
  independent method}, AIP Conf. Proc. 904~(1) (2007) 257--268.
\newblock \href {http://arxiv.org/abs/nucl-ex/0703031}
  {\path{arXiv:nucl-ex/0703031}}, \href {https://doi.org/10.1063/1.2734310}
  {\path{doi:10.1063/1.2734310}}.

\bibitem{Papanicolas:2012}
C.~N. Papanicolas, E.~Stiliaris, {A novel method of data analysis for hadronic
  physics}, arXiv 1205.6505 (5 2012).
\newblock \href {http://arxiv.org/abs/1205.6505} {\path{arXiv:1205.6505}},
  \href {https://doi.org/10.48550/ARXIV.1205.6505}
  {\path{doi:10.48550/ARXIV.1205.6505}}.

\bibitem{PCAStackExchange}
amoeba (https://stats.stackexchange.com/users/28666/amoeba),
  \href{https://stats.stackexchange.com/q/115477}{How to perform
  cross-validation for {PCA} to determine the number of principal components?},
  Cross Validated (2018).
\newline\urlprefix\url{https://stats.stackexchange.com/q/115477}

\end{thebibliography}


\begin{thebibliography}{1}
\expandafter\ifx\csname url\endcsname\relax
  \def\url#1{\texttt{#1}}\fi
\expandafter\ifx\csname urlprefix\endcsname\relax\def\urlprefix{URL }\fi
\expandafter\ifx\csname href\endcsname\relax
  \def\href#1#2{#2} \def\path#1{#1}\fi

\bibitem{Wosiek:1990}
B.~Wosiek, {Intermittency analysis of correlated data}, Acta Phys. Polon. B 21
  (1990) 1021--1030.

\bibitem{Michael:1994}
C.~Michael, {Fitting correlated data}, Phys. Rev. D 49 (1994) 2616--2619.
\newblock \href {http://arxiv.org/abs/hep-lat/9310026}
  {\path{arXiv:hep-lat/9310026}}, \href
  {https://doi.org/10.1103/PhysRevD.49.2616}
  {\path{doi:10.1103/PhysRevD.49.2616}}.

\end{thebibliography}

%% Appendices

\clearpage
\begin{appendices}
\renewcommand{\thesection}{\Alph{section}.}

\section{Determining the optimal number of\hfill\break components in PCA analysis}
\label{ap:optimal_PCs}

Performing PCA requires selecting a number of principal components (PCs) to be kept that is in a sense ``optimal"; care must be taken to make a proper choice: keeping too few PCs loses information in the samples, whereas keeping too many risks overfitting the specific samples used to determine PCA rotation.

A good metric for optimizing the number of PCs is the \emph{leave-one-out loss} criterion~\cite{PCAStackExchange}. The metric quantifies how well the original set of moments in $M$-space is reconstructed, when a given number of PCs are kept. Each of the $M$-bins in a $\Delta F_2(M)$ sample is in turn left out, and the remaining points are used to ``predict'' its value, via the direct and (pseudo-)inverse PCA transform. The squared differences between true \& predicted $\Delta F_2(M)$, summed over all $M$-bins and averaged over samples, constitute the leave-one-out loss score; minimizing this score over all options of \#PCs kept indicates the optimal number of components.

\begin{figure}
    \centering
    \includegraphics[width=.75\textwidth]{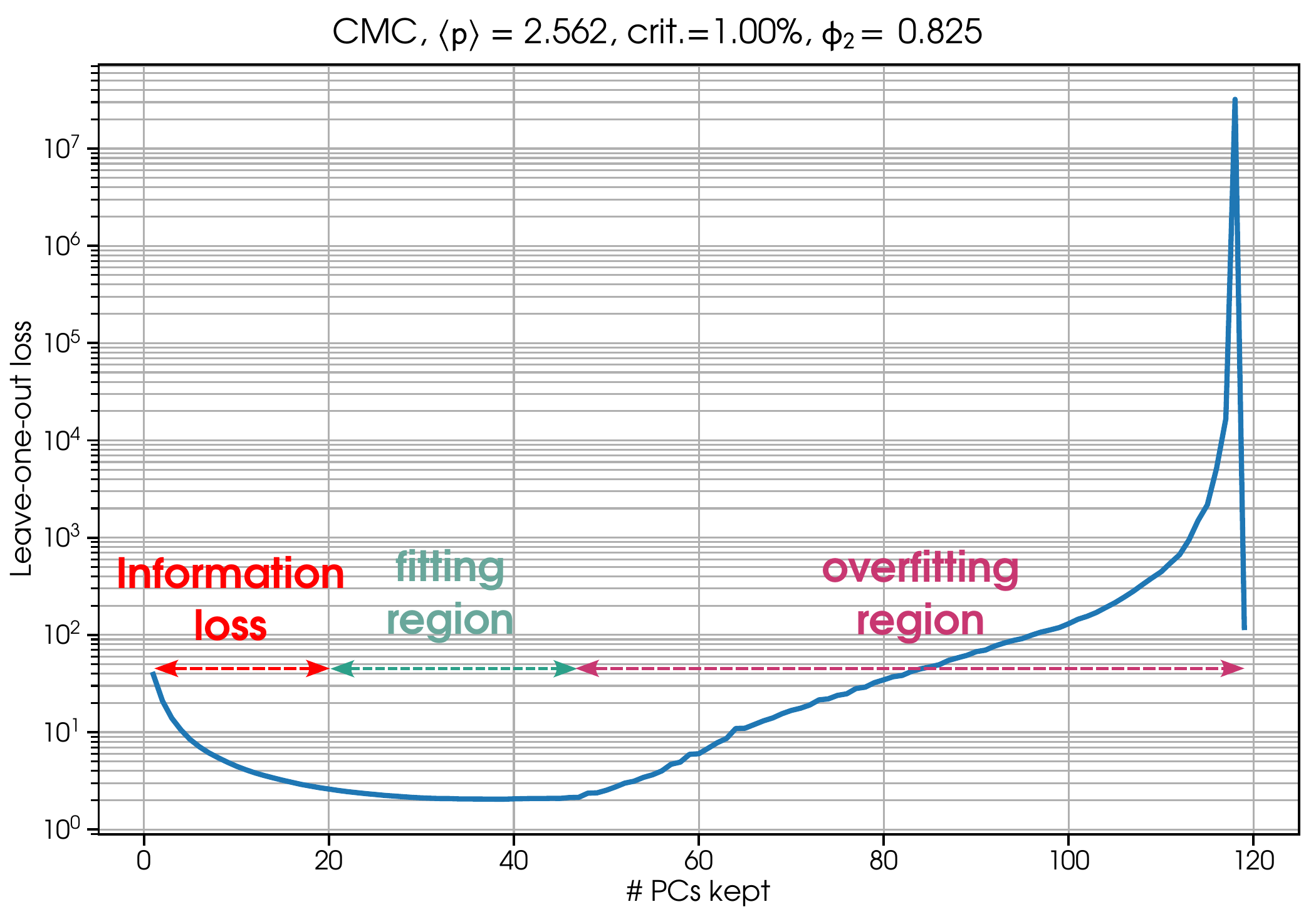}
    \caption{Leave-one-out loss mean test score as a function of \# PCs kept for the synthetic CMC dataset shown in Fig.~\ref{fig:PCA_application}.}
    \label{fig:leave-one-out-loss}
\end{figure}

Fig.~\ref{fig:leave-one-out-loss} shows the results of a cross-validation scan of leave-one-out test scores over the $\Delta F_2(M)$ bootstrap samples, for the synthetic CMC dataset shown in Fig.~\ref{fig:PCA_application}. From 1 up to 119 most significant PCs may be kept, in the $M$-range used in the analysis; the \# of PCs kept that minimizes leave-one-out loss is shown to be 38, although we observe a very wide and shallow valley, in the range $\sim 25-45$ PCs, where essentially the leave-one-out loss score is roughly constant, and minimal. In the analysis shown in Section~\ref{sec:PCA}, we have kept 38 components throughout.

\section{A $\chi^2$ residuals distribution test for the\break presence of higher order correlations in the PCs}
\label{ap:chi2_independence}

\begin{figure}[htb]
 \centering%
 \includegraphics[width=0.325\textheight]{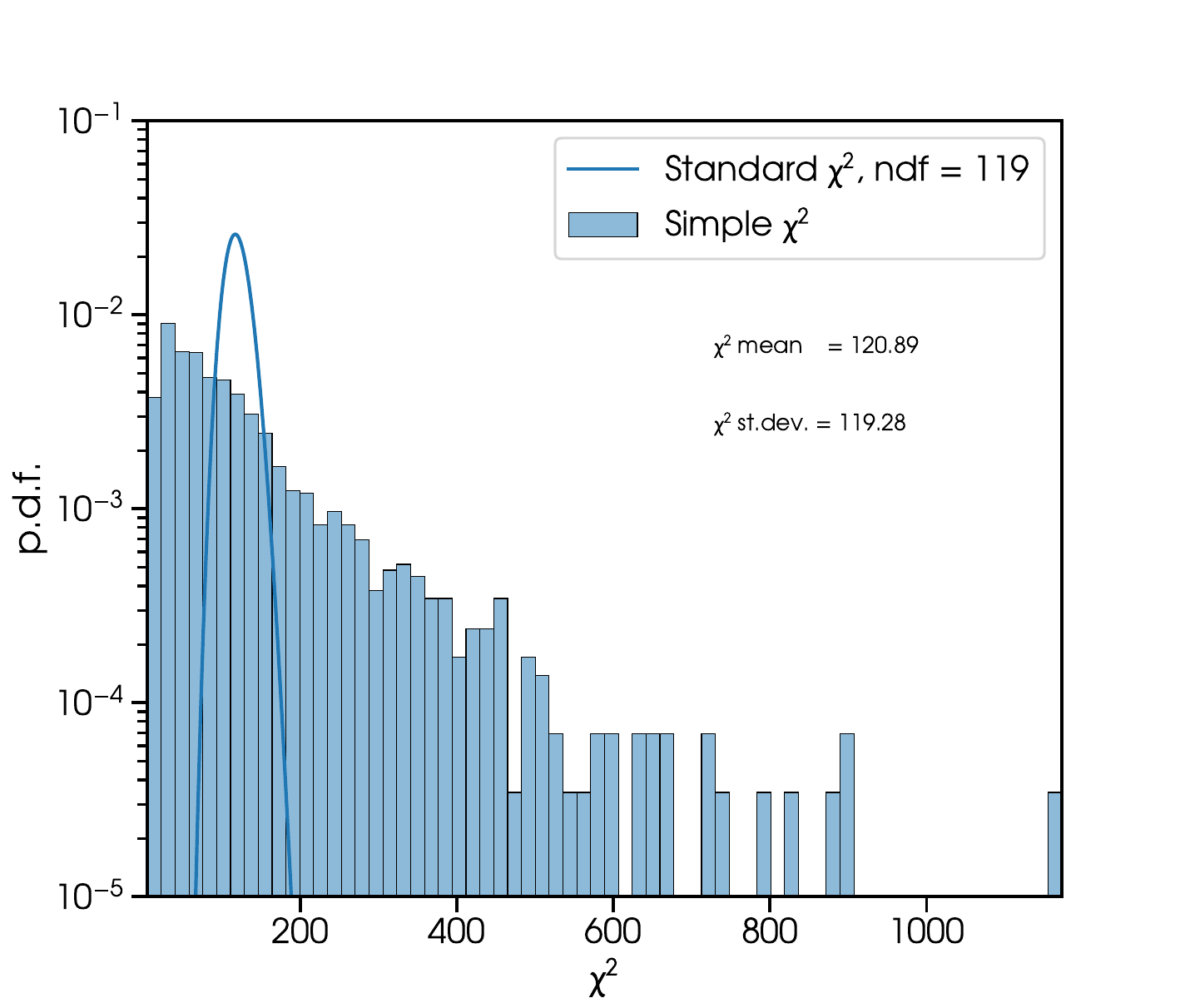}%
 \includegraphics[width=0.35\textheight]{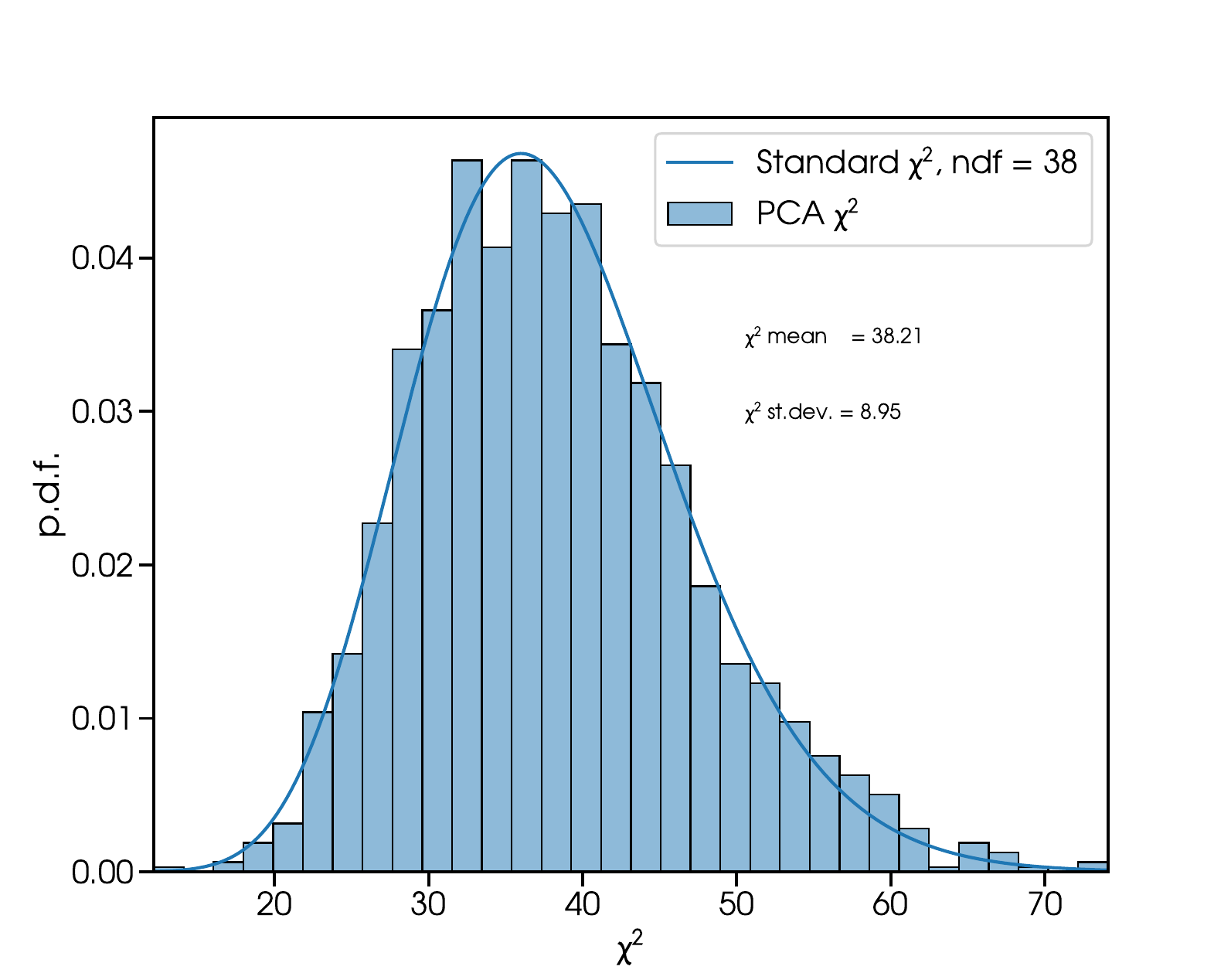}%
 \caption{\emph{Left:} Histogrammed distribution of the sum of squared residuals ($\chi^2$) for $\sim1$K bootstrap samples vs the original sample of the synthetic CMC data set appearing in Fig.~\ref{fig:PCA_application} \emph{(top left)}, in the original variables; the light blue curve illustrates the standard $\chi^2$ distribution with the corresponding number of degrees of freedom
($ndf = 119$). \emph{Right:} the corresponding $\chi^2$ distribution for the transformed PC variables, Fig.~\ref{fig:PCA_application} \emph{(bottom left)}, (\#PCs kept = 38) vs the  standard $\chi^2$ distribution with $ndf = 38$.}
 \label{fig:chi2_independence}
\end{figure}

In performing a $\chi^2$ test of a model to observed data, we implicitly assume that the sum of squared residuals between the model and data values, i.e. the $\chi^2$ statistic, follows a standard $\chi^2$ distribution with the appropriate number of degrees of freedom (ndf). For that to be true, the residuals must be random variables that are (i)
independent of each other, (ii) normally distributed, with (iii) zero mean (unbiased) and (iv) their distribution has a common variance $\sigma^2$; condition (iv) can be
relaxed, if the original data are normalized to unit variance, at the same time adjusting the number of degrees of freedom accordingly, if any parameters are estimated from the data.

We can empirically test assumptions (i-ii) by plotting the $\chi^2$ distribution (the sum of squared residuals between model and data), either for a large number of synthetic
simulated samples of a particular model vs the model trend (mean), or, in the case of experimental data, for a large number of bootstrap samples vs the original experimental sample. We then compare these distributions to the standard
$\chi^2$-distribution with the appropriate degrees of freedom. If variables are statistically independent and variation is (approximately) normal, i.e. gaussian, the standard $\chi^2$-distribution should be a good match to the empirically observed $\chi^2$ distribution. This amounts to a self-consistency test where variations are compared to the model where they originated from, by construction.

Fig.~\ref{fig:chi2_independence} shows the $\chi^2$ distribution for the residuals of $\sim 1$K bootstrap samples of the CMC data set, Fig.~\ref{fig:PCA_application}, both for the original variables \emph{(left)} as well as the 38 most significant PCs \emph{(right)}. We observe that the $\chi^2$ distribution for the original variables bears no resemblance to the standard $\chi^2$-distribution for the corresponding degrees of freedom –- or for that matter, for any
number of degrees of freedom. The empirical distribution is characterized by a very long tail, and while its mean $\chi^2$ value is close to that of the standard $\chi^2$-distribution, it has a large standard deviation comparable to the mean, resulting in a large spread of possible values for candidate samples. By contrast, the empirical $\chi^2$ distribution in the PCs is in excellent agreement with the standard $\chi^2$-distribution with the corresponding ndf.

In conclusion: in the case of the synthetic CMC data set shown in Fig.~\ref{fig:PCA_application}, empirical evidence from the $\chi^2$-distribution of bootstrap sample residuals justifies the assumption of negligible higher order correlations among the PCs; moreover, the PC transformation automatically eliminates \emph{linear} order correlations among PCs, leaving the PCs as effectively independent variables on which the PCA methodology, Section~\ref{sec:PCA}, can be safely applied.

Application of this test is highly recommended before applying the PCA methodology outlined in Section~\ref{sec:PCA} to any real-world set of factorial moment data.

\end{appendices}

\end{document}

\title{Supplemental materials to the manuscript ``A new approach to handling factorial moment correlations through principal component analysis''}

\author[1]{N.~Davis\footnote{email: nikolaos.davis@ifj.edu.pl}}
\affil[1]{Institute of Nuclear Physics PAN, ul. Radzikowskiego 152, 31-342, Krak\'ow, Ma\l{}opolska, Poland}
\date{}

\maketitle

\section{Comparison of correlated $\chi^2$ metric to simple $\chi^2$ in the rotated PCs via Principal Component Analysis (PCA)}
\label{sec:comparison_corchi2_pca}

Correlated $\chi^2$ fitting was introduced as a method for intermittency analysis in the early 90s~\cite{Wosiek:1990}, but has not been widely adopted since. The reason for this is the numerical instability of correlated $\chi^2$ fitting when the number of fitted points (bins $M$) is large, and the number of (bootstrap or independent) samples used to estimate the covariance matrix is relatively small. The essence of the problem is that the smallest eigenvalues of the (empirically estimated) covariance matrix will destabilize the fit, producing erroneous values for the fit parameters and their uncertainties, as shown in Ref.~\cite{Michael:1994}. Faced with these practical problems, the traditional approach was to perform a simple $\chi^2$, i.e. ignore bin correlations altogether. Only recently have better alternatives been employed, namely PCA, which allow one to take correlations into account in a numerically stable manner.

The rule of thumb proposed in Ref.~\cite{Michael:1994} is not to use correlated $\chi^2$ unless

\begin{equation}
 N > \max\left( D^2, 10 (D + 1) \right),
 \label{eq:rule_of_thumb}
\end{equation}

where $N$ is the number of data samples and $D$ is the number of points to be fitted. In the case of intermittency, $D$ corresponds to the total number of $M$-bins (per dimension subdivisions of transverse momentum space) studied ($M=119$ in the synthetic data set shown in Fig.~3 of the manuscript, discarding bins up to $M^2 = 10^3$), whereas $N$ is the number of bootstrap (or CMC) samples generated. Following the rule of thumb, Eq.~(\ref{eq:rule_of_thumb}), $D=119$ bins would require of the order of $\sim14$K samples for a reliable fit; in practice, $\sim1-10$K bootstrap samples can be generated before the calculation becomes prohibitively expensive, and even inflating that number despite the computational cost risks biasing the results, since bootstrap samples are not independent of each other and the original sample. To make matters worse, in the case of intermittency it is not reasonable to use an uncorrelated $\chi^2$ fit, as evidenced by the form of the $\chi^2$ residuals distribution, Appendix B of the manuscript.

In light of the above, the PCA can be viewed as a stabilizing technique for calculating the correlated $\chi^2$; as the eigenvalues of the covariance matrix are sorted from largest to smallest, and only a limited number of the largest eigenvalues are kept, the small, destabilizing eigenvalues do not enter the calculation at all. Combined with the use of a proper metric for selecting the optimal number of components (such as described in Appendix A of the manuscript), we obtain a reasonable approximation to the exact covariance matrix, and thus an accurate $\chi^2$ value of the model vs data.

\begin{figure}
 \centering\includegraphics[height=.271\textheight]{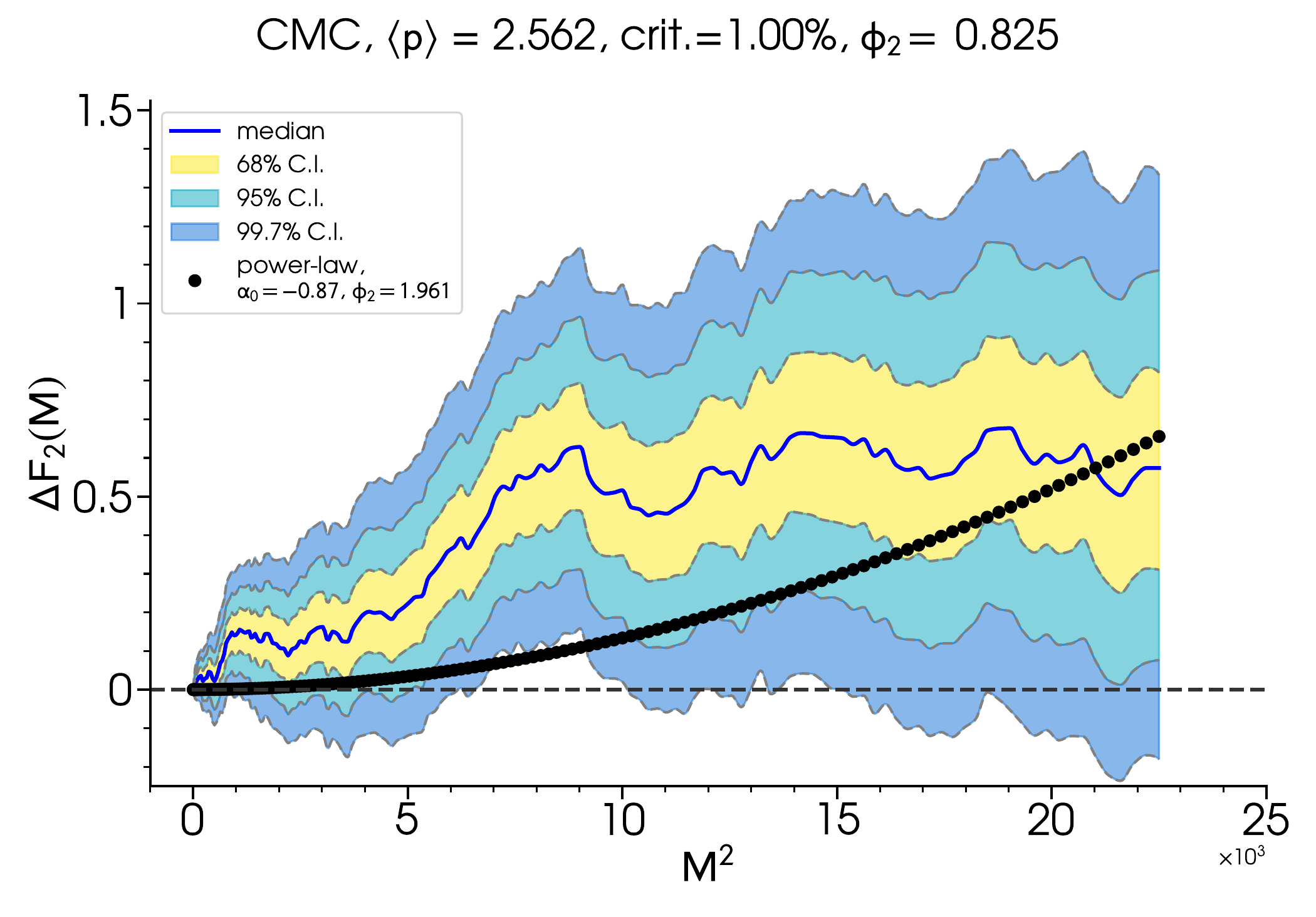}%
 \includegraphics[height=.271\textheight]{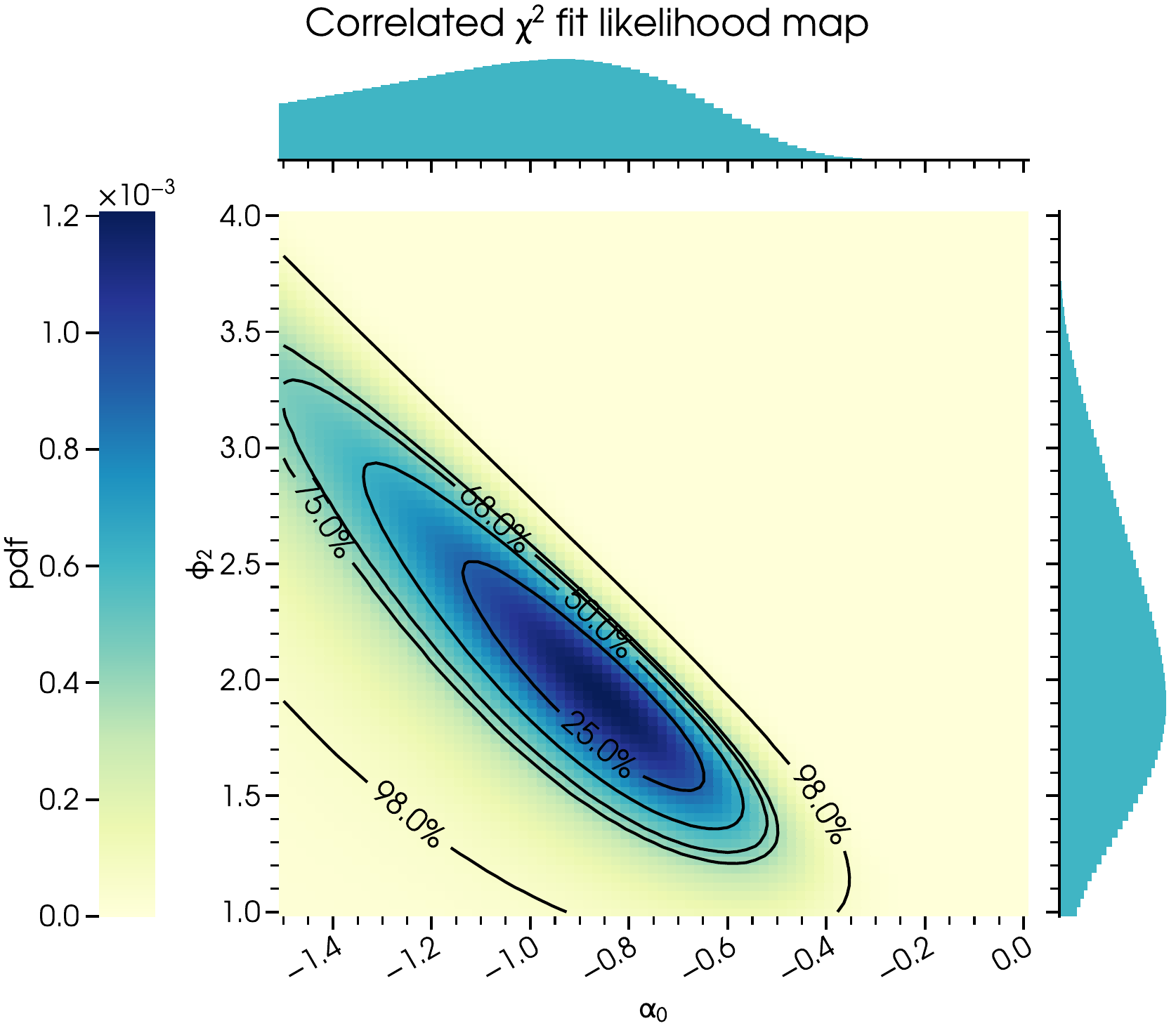}
 \caption{\emph{Left:} Correlator $\Delta{}F_2(M)$ of $\sim8$K bootstrap samples of the synthetic CMC dataset appearing in Fig.~3 (top left) of the manuscript, in the original variables; the correlated $\chi^2$ best fit power law as a function of $(\alpha_0, \varphi_2)$ is overlaid on the plot. Right: Normalized likelihood $\exp(-\chi^2/2) \Big/ \sum \exp(-\chi^2/2)$ for a grid of power-law parameters vs the synthetic CMC dataset, based on the correlated $\chi^2$. Marginal $\alpha_0$ $(\varphi_2)$ distribution plotted on the top (right) of 2D-plot.}
 \label{fig:methods_comparison_corfit}
\end{figure}

\begin{figure}
 \centering\includegraphics[height=.271\textheight]{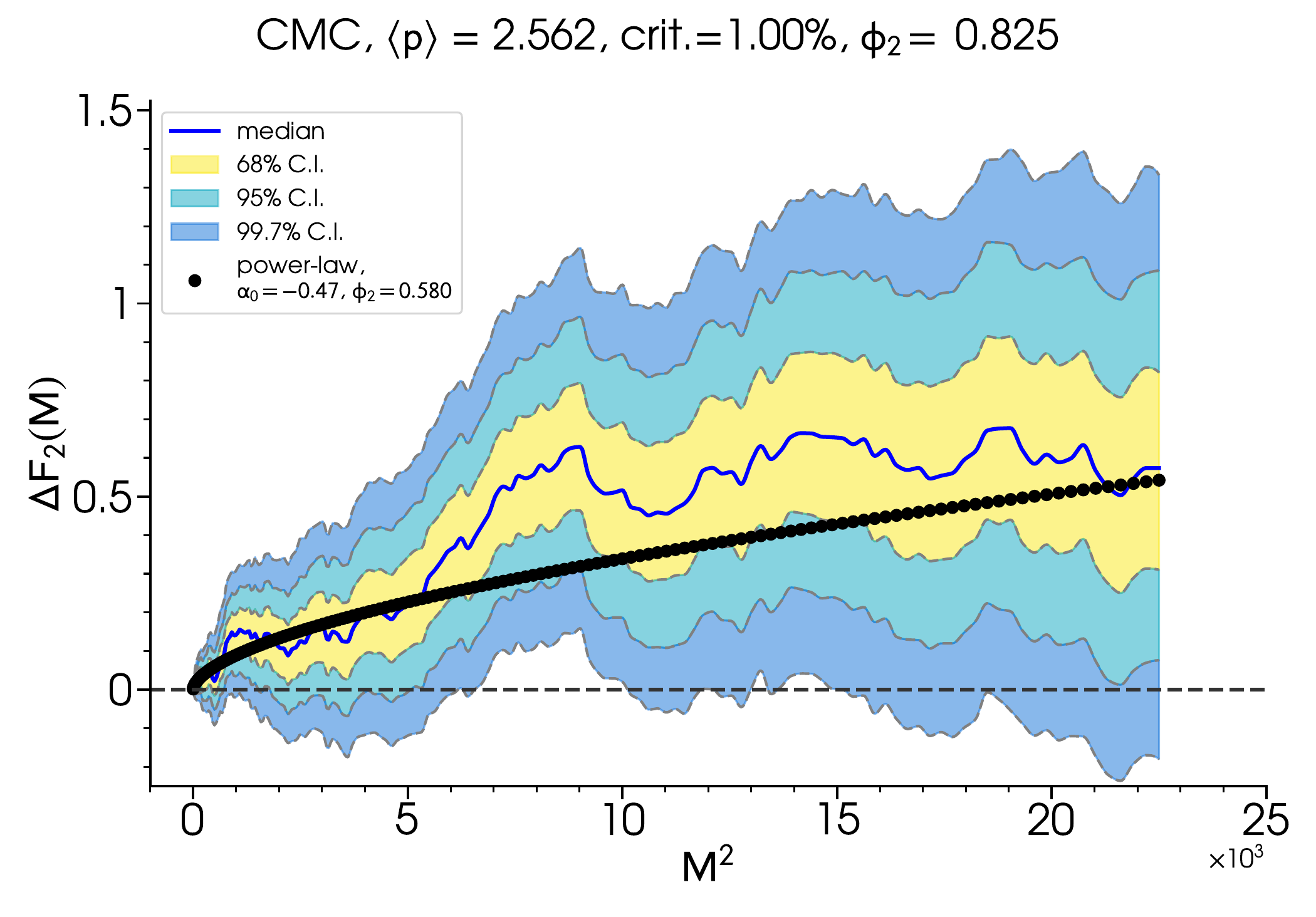}%
 \includegraphics[height=.271\textheight]{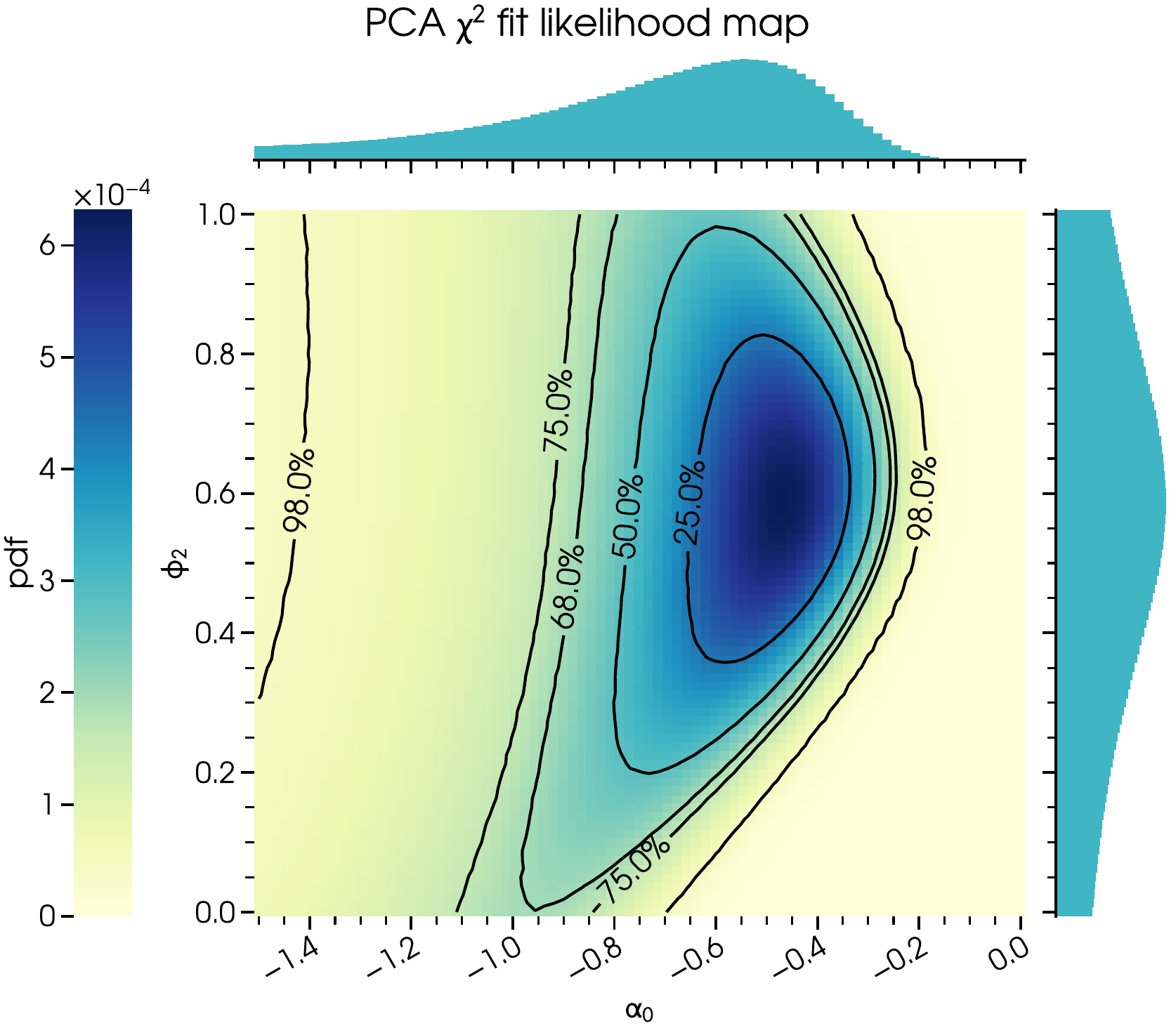}
 \caption{\emph{Left:} Correlator $\Delta{}F_2(M)$ of $\sim8$K bootstrap samples of the synthetic CMC dataset appearing in Fig.~3 (top left) of the manuscript, in the original variables; the PCA $\chi^2$ best fit power law as a function of $(\alpha_0, \varphi_2)$ is overlaid on the plot. Right: Normalized likelihood $\exp(-\chi^2/2) \Big/ \sum \exp(-\chi^2/2)$ for a grid of power-law parameters vs the synthetic CMC dataset, based on the PCA $\chi^2$. Marginal $\alpha_0$ $(\varphi_2)$ distribution plotted on the top (right) of 2D-plot.}
 \label{fig:methods_comparison_PCAfit}
\end{figure}

In Figs.~\ref{fig:methods_comparison_corfit}~\&~\ref{fig:methods_comparison_PCAfit}, we illustrate the difference in stability between the correlated $\chi^2$ metric and simple $\chi^2$ in principal components, using the synthetic CMC data set plus its bootstrap variations. Fig.~\ref{fig:methods_comparison_corfit}~\emph{(left)} shows $\Delta{}F_2$ for $\sim8$K bootstrap samples of the synthetic data set, with color bands corresponding to confidence intervals of the bootstrap distribution. The black points correspond to the best power-law model fit, i.e. the one that minimizes correlated $\chi^2$ (the $\Delta{}F_2$ bootstrap distribution is used to estimate the covariance matrix, which in turn defines the correlated $\chi^2$). Fig.~\ref{fig:methods_comparison_corfit}~\emph{(right)} shows the normalized likelihood (distribution) of a grid of power-law models when the correlated $\chi^2$ is used as a metric. This figure is analogous to Fig.~4 of the manuscript, except $\chi^2$ is calculated using the residuals in the original $M$-bins, and the full (bootstrap) covariance matrix is used, instead of a diagonal variance matrix as would in uncorrelated $\chi^2$. The contours enclose a certain percentage of all probability in the grid, indicated by the contour labels. We observe that the most likely model (equivalently, the model minimizing correlated $\chi^2$) sits at $\varphi_2 \sim 1.96$, a value that is unphysical, as $\varphi_2$ ranges in the $[0,1]$ interval. Furthermore, the contours are elongated, giving a very wide margin of error for the best fit values of power-law parameters.

In contrast, Fig.~\ref{fig:methods_comparison_PCAfit} shows the best fit results for the power-law when simple $\chi^2$ is used in the principal components (PCs) coordinates keeping the 38 most significant principal components. Now, the best fit value of $\varphi_2 \sim 0.58$ lies in the physical range $[0,1]$, and the 50\% contour, Fig.~\ref{fig:methods_comparison_PCAfit}~\emph{(right)}, contains the plug-in (``true'') value $\varphi_2 = 0.825$ used to obtain the synthetic data set. Furthermore, the contours are much more restricted in their range, and the power-law model parameters $(\alpha_0, \varphi_2)$ are almost independent of each other, as evidenced by the contours in Fig.~\ref{fig:methods_comparison_PCAfit}~\emph{(right)} approximately aligning with the $x/y$ axes.

\begin{figure}
 \centering\includegraphics[height=.271\textheight]{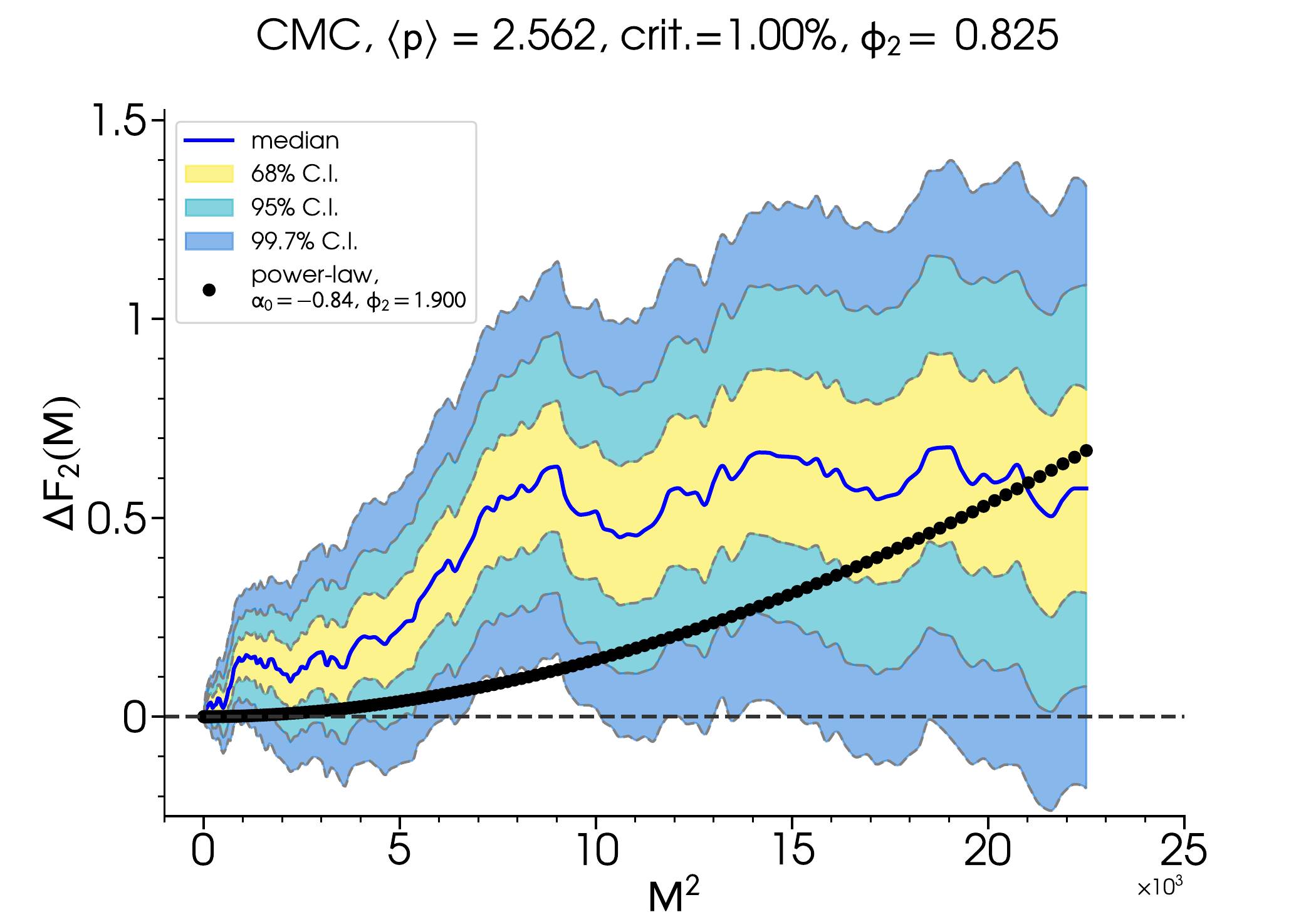}%
 \includegraphics[height=.271\textheight]{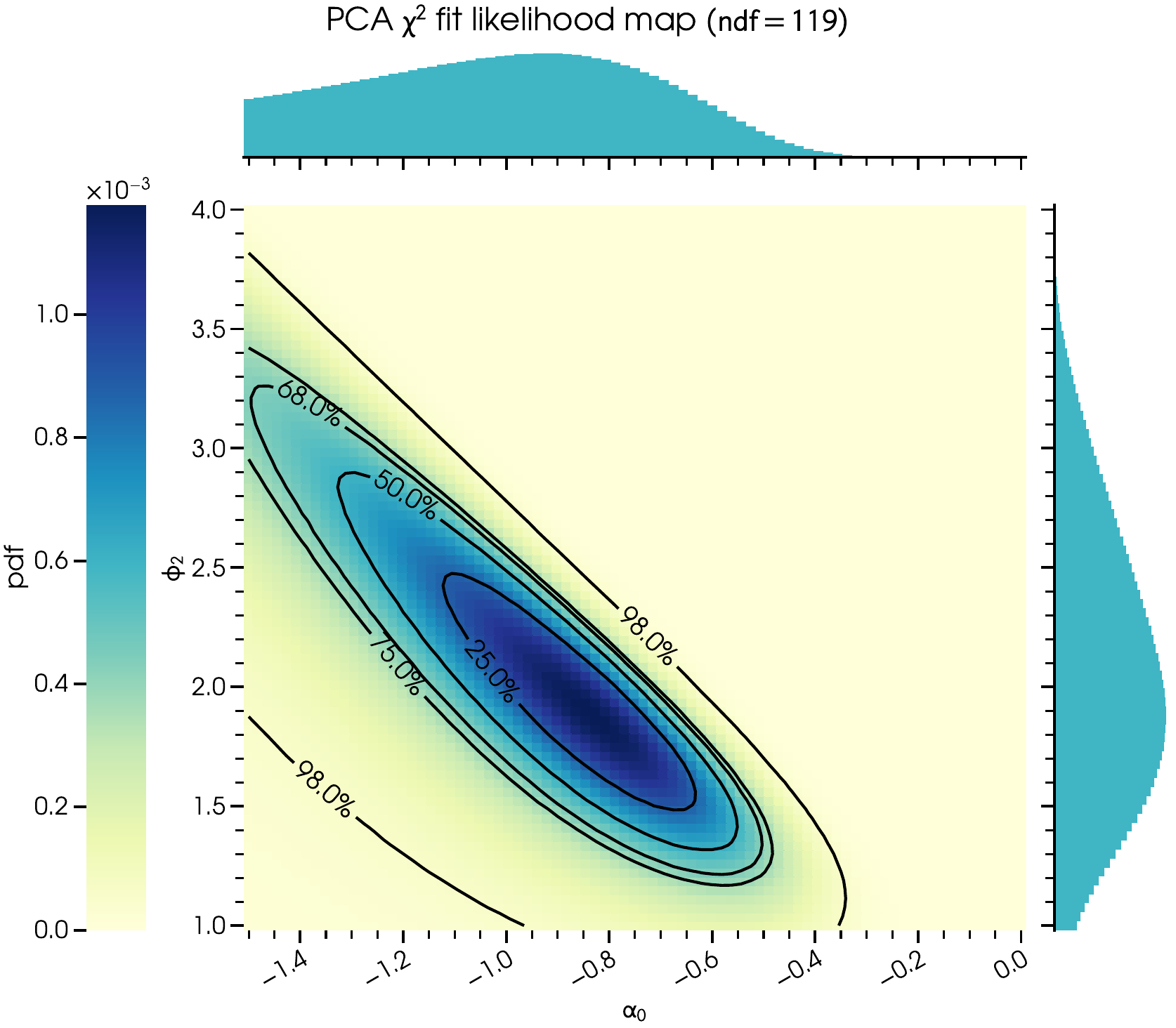}
 \caption{\emph{Left:} Correlator $\Delta{}F_2(M)$ of $\sim8$K bootstrap samples of the synthetic CMC dataset appearing in Fig.~3 (top left) of the manuscript, in the original variables; the PCA $\chi^2$ best fit power law as a function of $(\alpha_0, \varphi_2)$ is overlaid on the plot, for \#PCs~$ = 119$. Right: Normalized likelihood $\exp(-\chi^2/2) \Big/ \sum \exp(-\chi^2/2)$ for a grid of power-law parameters vs the synthetic CMC dataset, based on the PCA $\chi^2$ with \#PCs~$ = 119$. Marginal $\alpha_0$ $(\varphi_2)$ distribution plotted on the top (right) of 2D-plot.}
 \label{fig:methods_comparison_PCAfit_119}
\end{figure}

For a final illustration of the stabilizing role of PCA on the covariance matrix eigenvalues, Fig.~\ref{fig:methods_comparison_PCAfit_119} shows the best fit results for the power-law when simple $\chi^2$ is used in the principal components (PCs) coordinates keeping \emph{all} 119 principal components. The fitted curve and likelihood map are essentially identical to Fig.~\ref{fig:methods_comparison_corfit} (the correlated fit), as using all the eigenvalues amounts to perfectly reconstructing the empirical covariance matrix of the experimental bootstrap samples. It is therefore in the weeding out of smaller eigenvalues that the benefit of the PCA lies; both the PCA and correlated $\chi^2$ fit diagonalize the covariance matrix and eliminate (linear) correlations between the newly defined coordinates (the principal components); the PCA additionally allows us to keep only the most significant eigenvalues of the covariance matrix, discarding the possibly noisy smaller ones. Coupled with a metric for selecting the optimal number of components, such as the leave-one-out loss criterion (Appendix A in the manuscript), we obtain a robust methodology for handling correlated data.

%% Bibliography
\bibliographystyle{elsarticle-num}
\bibliography{my-bib-database}

%% Appendices

\clearpage
\begin{appendices}
\renewcommand{\thesection}{\Alph{section}.}

\end{appendices}